# A new twist on PIFE: photoisomerisation-related fluorescence enhancement

Evelyn Ploetz[a], Benjamin Ambrose[b], Anders Barth[c], Richard Börner[d], Felix Erichson[d], Achillefs N. Kapanidis[e], Harold D. Kim[f], Marcia Levitus[g], Timothy M. Lohman[h], Abhishek Mazumder[e], David S. Rueda[b], Fabio D. Steffen[i], Thorben Cordes[j]*, Steven W. Magennis[k]* and Eitan Lerner[l]*

[a] Department of Chemistry and Center for NanoScience (CeNS), Ludwig-Maximilians-Universität München, Butenandtstr. 5-13, 81377 München, Germany

[b] Department of Infectious Disease, Faculty of Medicine, Imperial College London, London, W12 0HS, UK
Single Molecule Imaging Group, MRC-London Institute of Medical Sciences, London, W12 0HS, UK

[c] Department of Bionanoscience, Kavli Institute of Nanoscience, Delft University of Technology, Delft 2629 HZ, The Netherlands

[d] Laserinstitut Hochschule Mittweida, Mittweida University of Applied Sciences, Mittweida, Germany

[e] Kavli Institute for Nanoscience Discovery, Department of Biological Physics, The University of Oxford, UK.

[f] School of Physics, Georgia Institute of Technology, 837 State Street, Atlanta, GA 30332, USA

[g] School of Molecular Sciences, Arizona State University, 551 E. University Drive, Tempe, AZ, 85287, USA.

[h] Department of Biochemistry and Molecular Biophysics, Washington University in St. Louis School of Medicine, St. Louis, MO 63110, USA

[i] Department of Chemistry, University of Zurich, Zurich, Switzerland

[j] Physical and Synthetic Biology, Faculty of Biology, Ludwig-Maximilians-Universität München, Großhadernerstr, 2-4, 82152 Planegg-Martinsried, Germany

[k] School of Chemistry, University of Glasgow, Joseph Black Building, University Avenue, Glasgow, G12 8QQ, UK

[l] Department of Biological Chemistry, Alexander Silberman Institute of Life Sciences, Faculty of Mathematics & Science, Edmond J. Safra Campus, Hebrew University of Jerusalem; Jerusalem 9190401, Israel
Center for Nanoscience and Nanotechnology, Hebrew University of Jerusalem; Jerusalem 9190401, Israel

*corresponding authors: cordes@bio.lmu.de, steven.magennis@glasgow.ac.uk, eitan.lerner@mail.huji.ac.il

**Abstract**
PIFE was first used as an acronym for protein-induced fluorescence enhancement, which refers to the increase in fluorescence observed upon the interaction of a fluorophore, such as a cyanine, with a protein. This fluorescence enhancement is due to changes in the rate of *cis*/*trans* photoisomerisation. It is clear now that this mechanism is generally applicable to interactions with any biomolecule and, in this review, we propose that PIFE is thereby renamed according to its fundamental working principle as photoisomerisation-related fluorescence enhancement, keeping the PIFE acronym intact. We discuss the photochemistry of cyanine fluorophores, the mechanism of PIFE, its advantages and limitations, and recent approaches to turn PIFE into a quantitative assay. We provide an overview of its current applications to different biomolecules and discuss potential future uses, including the study of protein-protein interactions, protein-ligand interactions and conformational changes in biomolecules.

## 1. Photoisomerisation as a modulator of fluorescence

Fluorescence spectroscopy is a powerful method for studying biological phenomena *in vitro* and *in vivo*. Fluorescent dyes are uniquely sensitive reporters of their immediate surroundings at different length- and time scales. For every need, there is likely a fluorescent reporter for the job [1–3]. Many fluorescence-based reporters and assays are based on fluorescence quenching, which can be either static by the formation of non-fluorescent complexes or dynamic by depopulation of the excited state. Quenching mechanisms include changes due to electron transfer, dye protonation, or excited-state photoisomerisation. The resulting changes in fluorescence intensity, lifetime, spectrum, or polarisation allow monitoring of the physicochemical condition in the vicinity of the reporter, including viscosity, pH, dye interactions, or the presence of ions and chemical groups. Combining two or more dyes, e.g., via Förster resonance energy transfer (FRET), provides additional capabilities to study interactions between the dye and quencher, either within a biomolecule of interest or between two biomolecules.

This review focuses on the principles and applications of spectroscopic and biophysical assays based on fluorescence modulation via photoisomerisation. These assays require the use of only one fluorophore (as opposed to FRET) but can still report on different ranges of interactions between a dye and a biomacromolecule. As we will show, these assays can be utilized in several inventive ways to provide structure- and species-specific information. A major inspiration for many recent developments was the pioneering work by Kozlov and Lohman [4] in 2002, in which the interactions of the *E. coli* single-strand binding protein (SSB) with fluorescein- and Cy3-labelled single-stranded DNA (ssDNA) were investigated using stopped-flow kinetics. In 2007, the Xie lab [5], and in 2009 the Ha lab [6] introduced the first single-molecule assays in which the same principle was used, i.e., to modulate fluorescence intensities by the binding/association of DNA-binding proteins in the vicinity of a cyanine fluorophore conjugated to DNA. The term *protein-induced fluorescence enhancement* (PIFE) was subsequently coined by Myong *et al.* [6], and PIFE is now used for various assays to study biomolecular interactions and structures as also summarized in previous reviews [7–9]. This review will start with a historical perspective of PIFE, followed by a discussion of the latest developments and possible future avenues with a focus on single-molecule applications. Finally, we propose changing the original PIFE acronym to **p**hoto**i**somerisation-related **f**luorescence **e**nhancement, which encompasses all related methods.

## 2. "PIFE: from old to new"
### 2.1 Photophysical background to PIFE

PIFE involves changes in the fluorescence quantum yield (QY), brightness and lifetime that are caused by distinct dye microenvironments. For cyanine fluorophores, which are typically used in PIFE assays, these changes are caused by competition between excited-state deactivation pathways that include radiative and non-radiative transitions to the electronic ground state in conjunction with a *cis-trans* isomerisation of the molecule. As a representative example, we consider a dye such as sulfo-Cy3 (sCy3) with a fluorescing all-*trans* isomer (0°; Figure 1A, *trans*) and non-fluorescent mono-*cis* isomer (180°; Figure 1A), which differ by rotation around θ (Figure 1A). The fluorescence QY values of the *cis* isomers of related tri- and pentamethine cyanines were estimated as to be ≤0.004 [10] and ≤0.01 [11], respectively. These values are significantly lower than those reported for rigidized versions of Cy3 (Cy3B, QY = 0.85 [12,13]) and Cy5 (Cy5B, QY = 0.69 [14], which are locked in the trans conformation. The mono-*cis* ground-state can form as a result of photoisomerisation, but thermally converts back into the more stable *trans* ground-state with rate $k_{gs}$ (Figure 1B) on the order of microseconds, and sometimes even milliseconds [15]. The observed emission of the dye (e.g., in a biophysical assay) is a result of the following processes: continuous excitation of the ground-state trans isomer at the appropriate excitation wavelength populates the excited state $S_T^*$ of the more stable *trans* isomer with rate $k_{ex,T}$. The latter can either decay to the *trans* ground state by internal conversion and fluorescence with the rate $k_T$, or photoisomerize via a 90°-twisted state (90°; Figure 1A/B, *twisted*) resulting in the formation of both the brighter *trans* and dimmer *cis* ground state isomers. The branching ratio between *cis* (at rate $k_{90\to C}$) or *trans* (at rate $k_{90\to T}$) isomers is governed by the position of the excited *twisted* state minimal energy and its maximal energy point at ground-state (Figure 1B). Importantly, any de-excitation pathway from out-of-plane excited-states, as well as the direct deactivation of the excited-state *cis* isomer $S_C^*$ via internal conversion with rate $k_C$, are always fully non-radiative [15,16] (Figure 1B). For the dyes Cy3 and Cy5, it is known that the ground-state *cis* isomer can be directly excited ($k_{ex,C}$) with red-shifted excitation relative to that of the *trans* isomer [15,17].

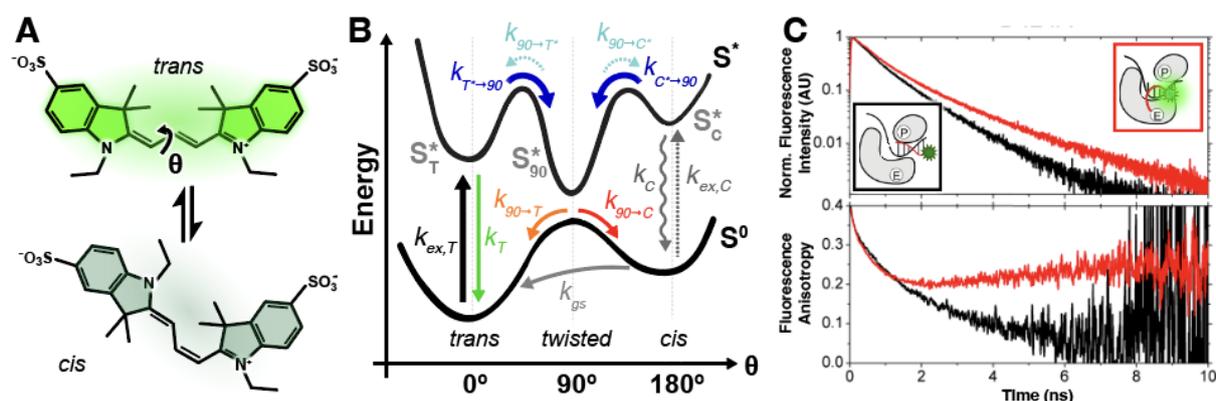

**Figure 1. PIFE concepts. (A)** Molecular structure of the cyanine dye sCy3 as *trans* (top) and *cis* isomer (bottom). Isomerization along the polymethine chain modulates the fluorescence of sCy3. **(B)** Energy diagram of sCy3 as a function of the rotation coordinate θ in the *trans* (0º) and *cis* (180º) state. Upon excitation into the excited *trans* state ($S_T^*$), deactivation occurs upon internal conversion and fluorescence (summarized by the decay rate $k_T$) or by isomerization ($k_{T^*\to 90}$) into the twisted state (90º). The excited *cis* state ($S_C^*$) decays via internal conversion to the *cis* ground state with a decay rate ($k_C$) or by isomerization ($k_{C^*\to 90}$). From the excited-state minimum in the twisted state, sCy3 forms the *trans* and *cis* ground state with rates $k_{90\to T}$ or $k_{90\to C}$, respectively. In the ground state, the reconversion from *cis* to *trans* isomer is again thermally driven with a rate $k_{gs}$. Adapted from Lerner, Ploetz, et al. [18] under the terms of the Creative Commons CC-BY License 4.0. **(C)** Using smPIFE at the single-molecule level allows, for example, monitoring the position of a Cy3-labelled dsDNA construct outside (black) and inside (red) a Klenow fragment via time-resolved fluorescence (top) and anisotropy (bottom). The transition of the primer to the exonuclease site pulls the Cy3-labelled fragment from a solvent-exposed to protein-surrounded position leading to a change in environment detected by PIFE. License: C) Adapted with permission from {Stennett E M S, Ciuba M A, Lin S and Levitus M 2015 Demystifying PIFE: The Photophysics Behind the Protein-Induced Fluorescence Enhancement Phenomenon in Cy3 The Journal of Physical Chemistry Letters 6 1819–23} [19]. Copyright {2015} American Chemical Society.

Overall, the observed brightness or fluorescence intensity is governed by the relative populations of 'brighter' *trans* and 'dimmer' *cis* isomers in the photodynamic equilibrium and the intrinsic non-radiative and radiative decay pathways in relation to photoisomerisation. Typically, the fluorescence lifetimes of dyes such as Cy3 are in the range of hundreds of picoseconds due to efficient photoisomerisation. Importantly, local viscosity and temperature impact this process and thus also excited state lifetimes [20–22], since rate $k_{T^*\to 90}$ is related to crossing an excited-state energy barrier (Figure 1B). Quantitative insights into these kinetics have been obtained by fluorescence correlation spectroscopy (FCS), providing estimates of isomerisation rates of Cy5 as a function of the irradiance, viscosity of the medium and temperature, and the effect of the conjugation to biomolecules [23].

The PIFE effect originates from a change in the local environment of the dye in terms of viscosity or specific interactions, which to a first approximation reduces the photoisomerisation rate $k_{T^*\to 90}$ (Figure 1B). This reduction increases the population of the excited *trans* isomer and decreases that of the excited *twisted* state and *cis* isomer, all of which increase the observed brightness and excited-state lifetime. Importantly, viscosity can vary due to solvents or co-solvents (e.g., high concentrations of viscogens and osmolytes) but also increase due to steric obstruction when the dye is conjugated to a biopolymer or its local environment changes due to biomolecular interactions. Such changes in steric obstruction (or microviscosity) of dyes are dubbed PIFE effects and have been used extensively to study biomolecular binding as well as local structural dynamics (Figure 1C).

**2.2 Origins of dye molecules for PIFE**

The idea of exploiting the photophysical properties of Cy3 and related cyanine dyes to investigate nucleic acid-protein interactions has a long history. Cyanine dyes are among the oldest and most investigated synthetic dyes. Studies in solution date back to the 1950s [24]. The formation of transient isomers from the singlet excited state and their subsequent reconversion to the stable form was recognized already in 1966 [25]. As shown in Figure 1, efficient internal conversion via photoisomerisation from the singlet excited state is responsible for the low fluorescence QY and short fluorescence lifetime of cyanine dyes such as Cy3 in solution [26–28]. The effect of solvent viscosity on these processes was thoroughly investigated in the 1980s and early 1990s using transient absorption and picosecond time-resolved spectroscopy [20,27,29–32]. These studies established a relationship between solvent viscosity and the rate of photoisomerisation, which ultimately governs the excited-state lifetime. Back then, cyanine dyes were primarily used as laser dyes, photoinitiators, and spectral sensitizers for silver halide photography and photodynamic therapy. Biological applications primarily used lipid-linked cyanines such as DiI (i.e., 1,1'-Dioctadecyl-3,3,3',3'-Tetramethylindocarbocyanine) or DiD (i.e., 1,1'-Dioctadecyl-3,3,3',3'-Tetramethylindodicarbocyanine) as membrane probes.

Alan Waggoner, a Professor of Biological Sciences at Carnegie Mellon University, recognised the potential of these fluorescent compounds as probes to visualise biochemical processes and cellular functions. His team designed new cyanine dyes with sulfonates coupled directly to the indolenine rings to prevent aggregation and improve water solubility (Figure 2). The series is now known as "Cy-dyes" and was synthesised and popularised as the NHS ester derivatives for labelling macromolecules [33]. Importantly, many more structural variants of the Cy-dyes emerged over the years, where better water solubility was achieved by varying the number of sulfonates on the indolenine rings (Figure 2). As highlighted earlier [7], the Cy-dyes lack a consistent nomenclature, and we suggest here indicating the degree of sulfonation for Cy-dyes in the name used (Figure 2) [34]: Cy for the unsulfonated version and sulfo-Cy (sCy) for the double sulfonated version, to clearly distinguish them from other members of the structural family, e.g., from the AF dyes and the Alexa Fluor series, and to relate a unique molecular structure to the name used. As a side note, we refer the reader to a recent work showing that dyes from the commercial AF series and from the Alexa Fluor series turn out to have different chemical structures as well as different photophysical features [34], hence the distinction between the two names (see examples in Fig. 2).

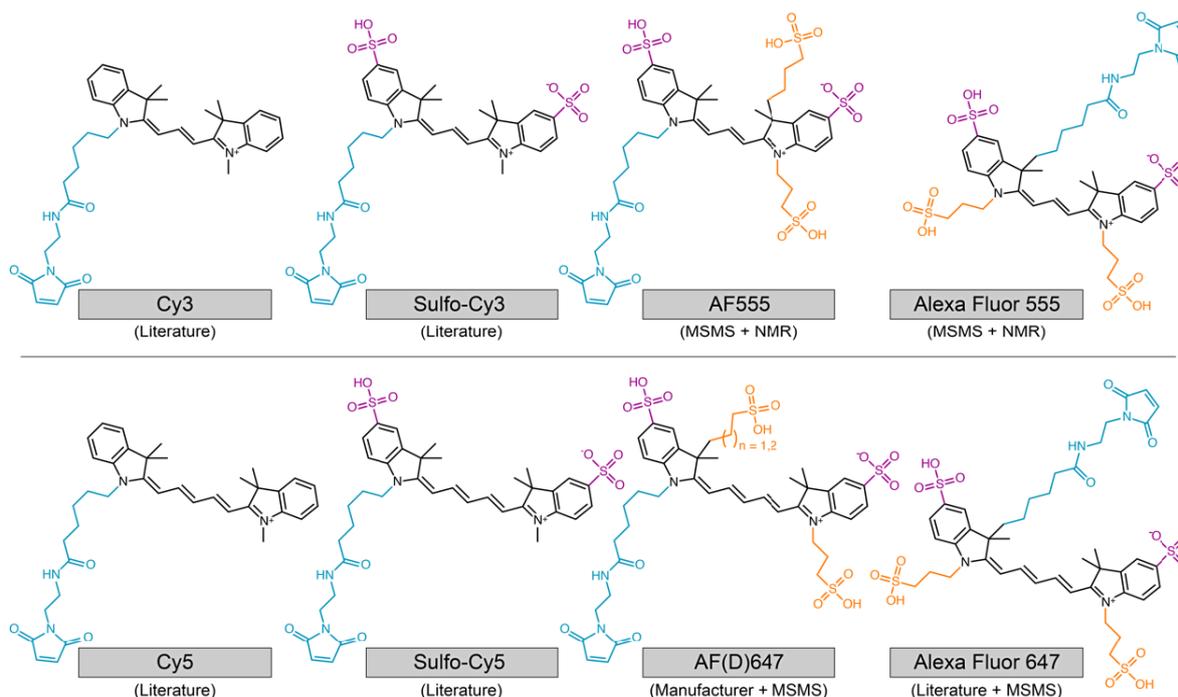

**Figure 2. Confirmed chemical structures of cyanine dyes (Cy-, Alexa Fluor-, and the AF-series) frequently used for biological applications.** Please note that the AF-dye homologs of Cy5 are available in two distinct versions called AFD647 (n = 1) and AF647 (n = 2). Formerly unpublished structures were confirmed by NMR and MS/MS [34]. To highlight structural differences compared to the parental cyanine fluorophore, we coloured linkers for labelling, e.g., via maleimide groups in blue, sulfo-groups in purple and sulfonated alkyl groups in orange. Please note that linkers for other type of dyes might differ in length.

The demand for cyanine-labelled oligonucleotides soared with the rise of single-cell flow cytometry, quantitative PCR, the development of single-molecule detection and imaging techniques, and many other important biotechnological applications (e.g., see [35–39]). Most of these applications benefit from bright labels with fluorescence properties that are insensitive to environmental changes, and in this context, the sensitivity of Cy3 to environmental conditions was initially perceived as a nuisance [12]. However, researchers quickly realised how to utilise this sensitivity to probe molecular interactions.

### 2.3 First applications of PIFE to protein-nucleic acid interactions

The focus of this review is on the photophysical basis and applications of photoisomerisation in various biochemical and biophysical assays. However, it is worth noting that this is certainly not the only approach for observing fluorescence modulation. Generally, the interactions of proteins with nucleic acids can be examined by labelling one of them (e.g., the nucleic acid) with an extrinsic dye that exhibits changes in fluorescence intensity or anisotropy in response to changes in its immediate microenvironment [40]. This alternative approach has been used extensively to study many interacting systems. In some cases, fluorescence enhancement occurs, while in other cases, fluorescence quenching is observed. Either effect can be used for monitoring interacting systems via fluorescence. One of the first ensemble-level studies using this approach examined the interaction of fluorescein-labelled tRNA with the elongation factor Tu, which was accompanied by fluorescence enhancement of fluorescein [41]. The binding of human β-DNA polymerase to a fluorescein-labelled ssDNA also exhibits fluorescence enhancement of fluorescein [42].

The first report of fluorescence enhancement involving photoisomerisation by Kozlov and Lohman [4], even before the PIFE acronym was coined, used fluorescein- and Cy3-labelled nucleic acids. This work reported a FRET-based measurement, which studied the interactions of the *E. coli* single-strand binding (SSB) protein with a Cy3-labeled ssDNA, (dT)$_{65}$, labelling its 3'-end with Cy3 and its 5'-end with Cy5 (Figure 3A). This length of ssDNA forms a 1:1 complex with the SSB tetramer, in which the DNA wraps around the tetramer such that the two ends of the ssDNA are brought in close proximity. Thus, an increase in FRET is expected upon SSB binding to the Cy3/Cy5 labelled $(dT)_{65}$. In fact, although the expected Cy5 fluorescence increase was observed, and a corresponding decrease in Cy3 fluorescence was expected, an unexpected Cy3 fluorescence increase was observed. Indeed, control experiments with $(dT)_{65}$ labelled solely with Cy3 also showed fluorescence enhancement. This increase was attributed to a direct interaction between Cy3 and the SSB protein that resulted in an increase in the Cy3 fluorescence QY.

This phenomenon was also observed during the interaction of the *E. coli* UvrD protein with a Cy3-labelled ssDNA. *E. coli* UvrD, as a monomer, is a rapid ATP-dependent translocase that moves along ssDNA in a 3'-to-5' direction [43]. An oligodeoxythymidylate ssDNA molecule, $(dT)_L$, of length *L*, was labelled with Cy3 at the 5'-end and was used for investigating the mechanism of UvrD monomer translocation (Figure 3B). UvrD will initially bind non-specifically, and thus randomly, to the $(dT)_L$-Cy3 molecules. Upon addition of ATP, the UvrD translocase will move along ssDNA (3' to 5') until it reaches the Cy3 label at the 5'-end, resulting in an enhancement of the Cy3 fluorescence intensity. Hence, one can measure the average time required for the enzyme to reach the 5'-end of the ssDNA. The kinetics of translocation can be examined in ensemble-level stopped-flow kinetic experiments using a series of $(dT)_L$-Cy3 molecules varying in length [43,44]. This approach has since been used to examine ssDNA translocation of a number of translocases and helicases, including *E. coli* Rep [45], *E. coli* RecBC [46,47], *E. coli* RecBCD [48], *B. stearothermophilus* PcrA [49], yeast Srs2 [50] and yeast Pif1 [51]. Lucius *et al.* [52] observed an interesting PIFE effect while monitoring DNA unwinding by RecBCD following a Cy3/Cy5 FRET signal. Just before the unwinding reaction was complete, resulting in DNA strand dissociation, a transient increase in the Cy5 fluorescence signal was observed. This is due to a Cy3 PIFE effect when RecBCD reaches the Cy3 dye that is then transferred via FRET to the Cy5 dye resulting in a transient Cy5 fluorescence increase before the expected Cy5 fluorescence decreases upon DNA strand separation. Other dyes, such as fluorescein and rhodamine red, can also be used in analogous experiments, although these dyes undergo fluorescence quenching upon interacting with UvrD [43].

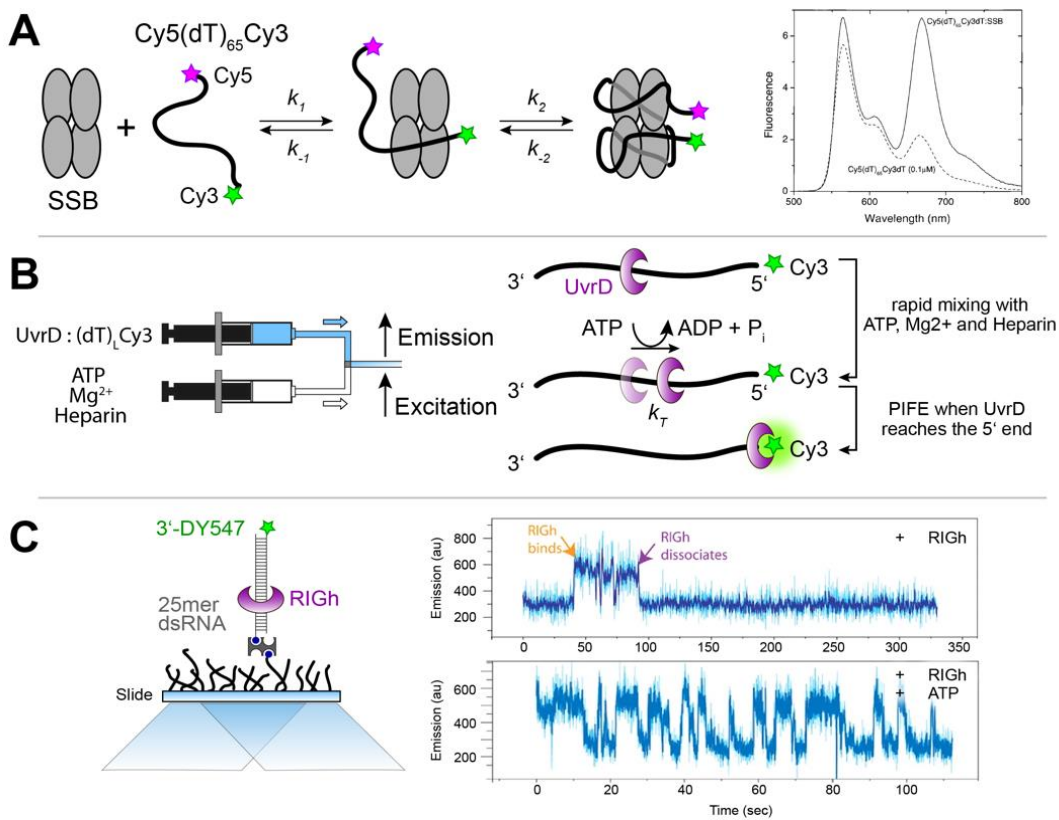

**Figure 3. Pioneering work using PIFE for probing the interaction between proteins and nucleic acids. (A)** The interaction between the tetrameric single-strand binding protein (SSB) and ssDNA leads to a 1:1 complex. Since the ssDNA is labelled with Cy5 and Cy3 at the 5' and 3' end, respectively, complex formation is observed by a shortening, leading to FRET between both dyes and an increase in brightness in Cy3 due to PIFE. **(B)** Translocation of the UvrD protein along ssDNA from the 3' to 5' end can be probed using stopped-flow experiments via the enhancement of the Cy3 fluorescence intensity once it reaches the 5' end. **(C)** Binding and translocation of the RIGh-I protein to single-stranded RNA was probed at the single-molecule level by TIRF microscopy and observed via fluorescence fluctuations in an ATP-dependent manner. Licenses: A) Adapted with permission from {Kozlov A G and Lohman T M 2002 Stopped-Flow Studies of the Kinetics of Single-Stranded DNA Binding and Wrapping around the Escherichia coli SSB Tetramer Biochemistry 41 6032–44} [4]. Copyright {2002} American Chemical Society; C) Adapted with permission from Reference [6].

In 2007, Luo *et al.* were the first to report the PIFE-based measurements and analysis of protein-dependent Cy3 intensity fluctuations at the single-molecule level [5]. This work is also significant because it was the first to provide a mechanistic explanation for the observed protein-dependent changes in Cy3's fluorescence intensity. The authors referred to the strong solvent viscosity dependence of the fluorescence QY of the dye and represented the potential energy surface commonly used to interpret solvent effects on photoisomerisation rates.

The term PIFE was coined in 2009 as an acronym for "*protein-induced fluorescence enhancement*" by Sua Myong, Taekjip Ha, and colleagues [6], reporting a single-molecule study of RIGh-I binding and its translocation on dsRNA. In this study, the authors used a dsRNA substrate, terminally labelled with DY547 (a dye closely related to Cy3), and observed fluctuations in the fluorescence intensity of the cyanine label that were associated with the repetitive binding and translocation of the protein on the dsRNA (Figure 3C). It was suggested that Cy3-based PIFE was a distance-dependent through-space phenomenon that could monitor short-distance changes (0-3 nm) [9,53]. However, it became clear that Cy3 fluorescence enhancement requires a direct interaction with the protein, such as that introduced by a steric obstruction, and it is not a through-space distance-dependent effect as it is in FRET [54]. Cy3-based PIFE signals have been used in various studies, including the measurement of diffusion along ssDNA of the human single-strand binding protein, RPA [55], DNA replication by the bacterial DNA Polymerase I Klenow Fragment [56,57], and the directional chemomechanical pushing of a protein along ssDNA by an ATP-dependent ssDNA translocase [58].

The photoisomerisation model to explain PIFE effects (Figure 1B) was later confirmed by a spectroscopic study from the Levitus lab using complexes of DNA and the DNA polymerase Klenow fragment [19]. This study demonstrated that the increase of the fluorescence QY and lifetime of Cy3 occurs in conditions where the dye is sterically constrained by the protein, as measured by a decrease in the rotational correlation time of the dye, and correlates with a decrease in the efficiency of photoisomerisation (Figure 1C). A negative correlation between the rotational correlation time of Cy3 and its fluorescence QY and lifetime was also observed in experiments with hRPA bound to ssDNA [54]. This study also established that the magnitude of the PIFE effect depends on the nature of the interactions between the protein and the Cy3 fluorophore, and varies when different regions of the protein interact with the dye.

It has also been demonstrated that Cy3 attached to DNA can undergo quenching, which was termed protein-induced fluorescence quenching (PIFQ), upon interaction with a protein, depending on the positioning of Cy3 within the DNA [59]. In that case, the quenching can alternatively be considered as a reduction in enhancement since it is due to a change in Cy3 from an already restricted state to a less restricted environment. Hence Cy3 can display either phenomenon depending on the context of the interaction [54,59,60].

## 3. New methods inspired by and related to PIFE

Since the early days, it has been clear that the modulation of photoisomerisation in various dyes can be used in several informative ways to report on a variety of different underlying features of the biomolecular sample. While in ensemble fluorescence measurements, PIFE was and is still used for probing biomolecular kinetics [4,52,61–66], applying PIFE to dyes that were bright enough using low background detectors paved the way for probing such features at the level of single biomolecules. Since the seminal works of the Levitus [13,19,67–69] and Myong [9,53,70] groups, single-molecule PIFE (smPIFE) has been used for studying protein-protein interactions, protein-nucleic acid interactions [18,56,71–74] in DNA and RNA structures, in both an inter- and intra-molecular fashion, and has been defined in many context-dependent ways with different acronyms (Figure 4). We discuss each of these in detail below.

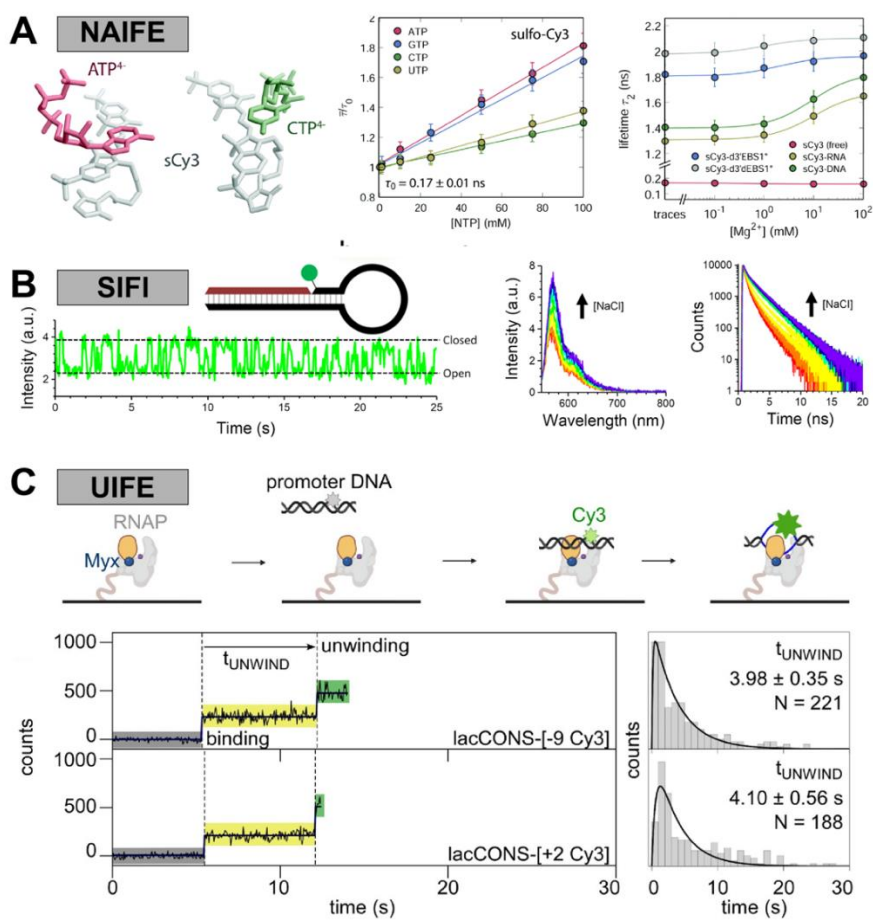

**Figure 4. Methods inspired by and related to PIFE. (A)** Nucleic acid-induced fluorescence enhancement (NAIFE). Interactions with nucleic acids lead to fluorescence enhancement of Cy3. **(B)** Stacking-induced fluorescence increase (SIFI). Stacking of Cy3 in a nick, gap or overhang of DNA leads to an increase in fluorescence intensity and lifetime. **(C)** Unwinding-induced fluorescence enhancement (UIFE). The unwinding of a dsDNA and bubble formation inside the bacterial RNA polymerase during transcription initiation can be investigated by labelling the nucleic acid with Cy3. Binding and melting of the DNA leads to contact between Cy3 and the RNAP and to an increase in fluorescence. Licenses: A) Reproduced from Ref. [75] with permission from the PCCP Owner Societies; B) Reproduced from Ref. [65] under the terms of a Creative Commons CC-BY 4.0 license. C) Reproduced from Mazumder et al., 2021, eLife [74] with permission, published under the Creative Commons Attribution 4.0 International Public License (CC BY 4.0; https://creativecommons.org/licenses/by/4.0/). Further reproduction of this panel would need permission from the copyright holder.

### 3.1 Nucleic acid-induced fluorescence enhancement (NAIFE)

Cyanines have become some of the most common fluorescence labels for conjugation to nucleic acids. One reason is their commercial availability in various functional forms, allowing chemical coupling to a base or the phosphate backbone [76–78]. The dyes can be incorporated into nucleic acids either during or after solid-phase synthesis or after *in vitro* transcription. Another reason for the popularity of cyanines is their photostability [79]. Interestingly, although free cyanine dyes in solution have low fluorescence QYs ($QY_{Cy3} = 0.1$) and extremely short fluorescence lifetimes ($\tau_F \leq 0.3$ ns for Cy3) and are prone to photo-destruction, they are considered to be photostable in the chemical environment of complex nucleic acid structures [80]. In the chemical environment of nucleic acids, Cy dyes experience more steric hindrance as compared to diffusing freely in solution, decreasing their photoisomerisation rate as described above (Figure 1). This effect can cause the fluorescence QY to increase up to QY=0.67, as was reported for Cy3 and its stiffened form, Cy3B ([Figure 6A](#)). Cy3B is chemically preventing *cis-trans* photoisomerization [12,81]. Similarly, the fluorescence lifetime increases by a factor of ~10, to 2.5 ns for Cy3B. The dependence of Cy dyes on the chemical environment of DNA was first described by Levitus and colleagues [13,67], showing that the modulation of the fluorescence QY and the lifetime of Cy3 [68] and Cy5 [82] depends on the DNA sequence.

The interaction of Cy dyes with nucleic acids can be divided into two categories: (i) charge-driven or electrostatic interactions and (ii) stacking interactions or hydrophobic effects. While the former describes the interaction of the dyes with the highly charged backbone of the nucleic acid sequence, the latter describes the interaction of the dyes with the ring system of the nucleobases. Both effects increase the photon yield and, thus, the molecular brightness of the Cy dyes. Collectively, this is referred to as nucleic acid-induced fluorescence enhancement (NAIFE). In the special case of ribonucleic acid, the effect is called RNA-induced fluorescence enhancement (RIFE).

(i) In the electrostatic interaction regime, the net charge difference leads to an attractive Coulomb force of the dye towards the nucleic acid backbone. This is of particular interest in the case of positively-charged fluorophores, like Cy3/Cy5, which tend to stack on the nucleic acid backbone. Sulfonated cyanine dyes such as sCy3/sCy5 carry negatively-charged sulfonate groups at neutral pH, which reduces dye-backbone interactions. The effect of the sulfonate groups is, in fact, twofold: on the one hand, the reduced interaction increases the mobility of the dye on the nucleic acid. Free dye rotation is a prerequisite for determining reliable distance information via FRET. On the other hand, reduced interaction increases the photoisomerisation probability, which, in return, leads to a decrease in brightness and causes the observed NAIFE/RIFE effects. Therefore, non-sulfonated Cy dyes are preferred for NAIFE and sulfonated sCy dyes for FRET experiments.

(ii) The interaction of the hydrophobic ring systems of the nucleobases and the Cy dye is characterised by entropy-driven stacking. This effect has been described for individual nucleotides, ssDNA and double-stranded DNA (dsDNA) labelled at the 3'/5'-end [83] and has been solved structurally by means of NMR [84]. Moreover, stacking can also occur with internally labelled nucleic acids. In the case of RNA, the interaction is strongly dependent on secondary structure elements and tertiary contacts. The formation of secondary elements is driven by monovalent metal ions such as K(I) and Na(I), whereas the formation of tertiary structures often depends on divalent metal ions such as Mg(II) and Ca(II) [85]. The more complex the surrounding structure of the dye is and the higher the binding affinity of divalent metal ions are, the more likely it is that dye-RNA interactions occur, thus increasing the NAIFE effect ([Figure 4A](#)). This was first demonstrated for sCy dyes by Steffen *et al.* in the presence of different RNA structures [75]. The dependence of NAIFE on the chemical microenvironment can now be used to investigate the degree of folding of the nucleic acid or the interaction probability with binding partners, such as hybridizing DNA fragments [86] or a recently developed DNA-aptamer sensor [87].

As absolute fluorescence intensity changes are experimentally susceptible to artifacts, (e.g., due to changes in the fluorescence background or the labelling efficiency of the host biomolecule) fluorescence lifetime-based measurements are an attractive alternative to exploit NAIFE (see also section 5.5 on lifetime-based smPIFE). The stronger the interaction between the dye and RNA or DNA, the lower the photoisomerisation rate and, thus, the longer the fluorescence lifetime. This association has been demonstrated by altering the complexity of the RNA chemical environment and the divalent metal ion-dependent binding of the dye to the RNA [75,88].

Analogous to fluorescence lifetime measurements, polarisation-resolved detection of the fluorescence signal yields another intensity-independent parameter, the dynamic fluorescence anisotropy *r*(*t*). Here, the rotational correlation time, $\tau_{\text{r,local}}$, and the residual anisotropy, $r_\infty$, of the dye are linked to NAIFE. Both parameters are sensitive to the local chemical environment of the dye and hence can describe the interaction with its immediate environment. These measurements can disentangle the interaction probability of Cy dyes with the host environment within the wobbling-in-cone model. The dynamic fluorescence anisotropy, $r(t)$, is divided into a local rotation of the dye within a cone [75,89] (see Eq. 1).

$$r_{\text{local}}(t) = (r_0 - r_\infty) \cdot e^{-t/\tau_{\text{r,local}}} + r_\infty \quad \text{(Eq. 1)}$$

and a stacked dye wobbling described by the global rotation correlation time, or simply put, the biomolecule tumbling time, of the host biomolecule [75] (see Eq. 2).

$$r_{\text{global}}(t) = r_{\text{local}} \cdot e^{-t/\tau_{\text{r, global}}} \quad \text{(Eq. 2)}$$

The fundamental anisotropy, $r_0$, is assumed to be independent of the chemical environment and is determined by the relative orientation of excitation and emission dipoles at time $t = 0$ s. The global rotation correlation time, $\tau_{\text{r, global}}$, might be determined from the hydrodynamic radius of the host molecule to further reduce the number of parameters in the fitting model.
The residual anisotropy accounts for a reduced depolarisation probability resulting from an energy barrier that prevents rotational diffusion of the fluorophore beyond a certain cone angle [90]. Steffen *et al.* showed that the cone angle is defined by the local chemical environment, hence by the surface of the host molecule. Thus, the local complexity of an RNA molecule perfectly correlates to an increase in $r_\infty$ [75].

Closely related to the parameters of the dynamic anisotropy are the so-called accessible volume (AV) and the contact volume (CV) of the dyes. The AV describes the entire explorable space of the dye determined by its geometric dimensions (linker length, dye radii) and the associated sterically-restricted molecular space [91,92]. The CV describes the volume where the dye and the host molecule interact with each other, for instance via stacking or hydrophobic interactions. The CV is included in the AV and has been introduced by Steffen *et al.* to provide a new model, termed the accessible contact volume (ACV) model, which distinguishes between freely-rotating and surface-stacked dye populations [75]. The ratio $\chi_{CV} = \frac{CV}{AV}$ has been shown to correlate to the local complexity of the host molecule. The experimental measure is a change in fluorescence lifetime and dynamic anisotropy. Here, $r_\infty$ characterises the insufficient depolarisation due to the local chemical environment of the dye and is a measure of the cone angle of the CV.

In summary, the more complex the chemical environment is, the larger the $\chi_{CV}$ becomes. The greater the CV is, the shorter $\tau_{\text{r,local}}$ becomes and hence the greater $r_\infty$ of the dye becomes. Similarly, the occupation of binding sites for divalent metal ions affects the surface stacking probability and is thus visible through NAIFE.

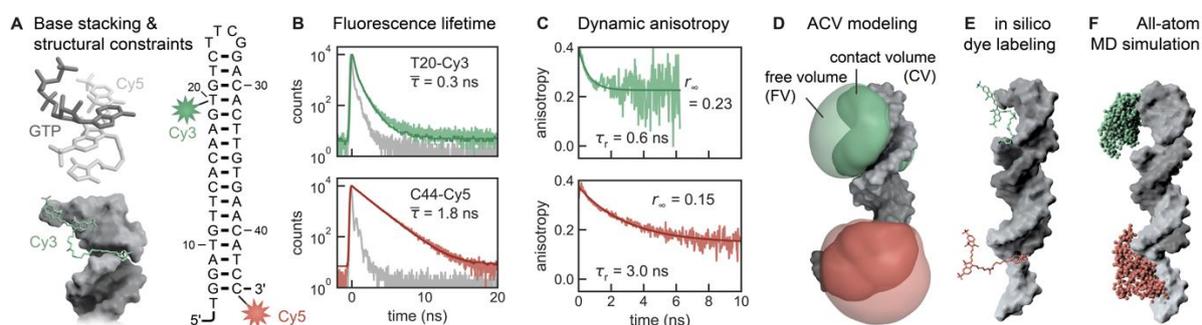

**Figure. 5. Photophysical measurements and computational modelling of NAIFE**. (A) Photoisomerisation of cyanine dyes is reduced by stacking on nucleobases and interaction with secondary and tertiary structure elements of RNA. (B) The average fluorescence lifetime, $\bar{\tau}$, is modulated depending on the degree of dye-RNA interaction. (C) The rotational correlation time, $\tau_r$, and the residual anisotropy, $r_\infty$, reflect the motional restriction of the dye by the RNA. (D) Surface trapping is modelled by the accessible contact volume (ACV). (E) An atomic-level description of dye-RNA contacts is provided by *in silico* labelling and by subsequent (F) molecular dynamics simulations.

### 3.2 Stacking-induced fluorescence increase (SIFI)

A DNA hairpin labelled with sCy3 on a 3′ dT (Figure 4B) exhibited increased fluorescence intensity that was associated with hairpin closing [93]. By examining a series of DNA hairpins and duplexes, it was shown that sCy3 undergoes site-specific stacking in a nick, gap or overhang region of duplex DNA. The sCy3 showed changes in fluorescence intensity at both the ensemble and single-molecule levels, and corresponding changes in fluorescence lifetimes were also observed at the ensemble level. The increase in fluorescence intensity or lifetime was attributed to a reduction in the rate of photoisomerisation upon stacking and hence was termed stacking-induced fluorescence increase (SIFI) [93,94]. This specific stacking interaction, and the previously reported stacking of cyanine dyes on the blunt end of duplex DNA [95,96], and on G-quadruplexes [97] should be considered as a subset of NAIFE (section 3.1).

Double labelling of a DNA hairpin with sCy3 and sCy5 as a FRET donor and acceptor, respectively, allowed a direct comparison of FRET and SIFI [94]. With both dyes fluorescently active, a FRET increase was observed upon hairpin closing, with sCy3 transitioning from high (open hairpin) to lower fluorescence intensity (closed hairpin). Following acceptor photobleaching, the sCy3 continued to exhibit intensity fluctuations but now transitioning from the same high fluorescence intensity as the FRET-active hairpin to an even higher intensity, which was due to the closing of the hairpin, stacking of the sCy3 on DNA and subsequently a reduction in photoisomerisation. Analysis of the two-state dynamics using hidden Markov modelling reported that the same opening and closing rates could be recovered via both FRET and SIFI. The ability to probe such global structural changes using only a single dye could be advantageous since it requires less synthetic modification, less chemical perturbation to the native behaviour and frees up a spectral window, which can be used for combining other fluorescence measurements.

It was also shown that fluorescence intensities and lifetimes of sCy3 are extremely sensitive to local changes at the site of stacking [93]. This was exploited by designing a DNA structure containing an abasic site in duplex DNA at distances of ≤20 nucleotides away from the sCy3 stacking site. The average fluorescence lifetime of the sCy3 was found to oscillate as a function of the distance from the abasic site; this was attributed to long-range, through-backbone allosteric interactions, which modulate the local sCy3 stacking interaction. This agreed with earlier studies of allostery in protein-DNA interactions, whereby the binding of one protein on one site in DNA affected the binding of a second protein on another site further along the duplex [98,99].

### 3.3 Unwinding-induced fluorescence enhancement (UIFE)

Environment-dependent fluorescence intensity enhancement of Cy3 has also been exploited to study the formation of an unwound transcription initiation bubble comprised of ssDNA segments by RNA polymerase (RNAP) as it binds and unwinds promoter dsDNA. In a first ensemble-level study, Ko and Heyduk [100] reported that the fluorescence intensity from a Cy3 strategically placed on promoter DNA showed a ~two-fold increase upon binding of RNAP. Subsequently, the Cy3 signature showed a similar decrease after transcription initiation and promoter escape. The results and control experiments described in the same report indicated that the observed fluorescence intensity increase is due to the unwinding of dsDNA to ssDNA upon RNAP binding, while a decrease results from the rewinding of ssDNA to dsDNA upon promoter escape. The large ~two-fold fluorescence enhancement in unwinding-induced fluorescence enhancement (UIFE) assays could possibly result from a combination of binding of RNAP to the promoter dsDNA, unwinding of promoter dsDNA to ssDNA segments, and subsequent conformational changes involving the unwound ssDNA segment and RNAP. The ensemble assay is simple and straightforward to implement and has been used extensively in studies investigating the mechanism of promoter unwinding and promoter escape in transcription by several groups [62,100–103].

Later, the Ha lab implemented a single-molecule UIFE (smUIFE) assay to study the kinetics and mechanism of transcription initiation by a phage T7 RNAP [104]. More recently, similar smUIFE experiments were used in real-time single-molecule assays investigating the promoter unwinding mechanism by a bacterial RNAP (Figure 4C). Here, the authors monitored the unwinding kinetics of the upstream and downstream segments of a promoter fragment to show that unwinding occurs in steps that proceed from upstream towards the downstream direction [74]. The smUIFE assays can potentially be combined with high-throughput single-molecule studies of large promoter sequence libraries, enabling a complete dissection of the promoter sequence dependence during this stage of transcription initiation. Similar assays can also be used in other processes that involve DNA unwinding and rewinding, such as replication initiation and nucleic acid helicase and topoisomerase activities. Notably, such assays will carry different signal contributions from the unwinding and the rewinding process, as well as from the proximity of the protein machinery.

Importantly, fluorescence enhancement mechanisms similar to the ones mentioned above exist, where stabilization of the planar excited-state occurs via binding to a molecular scaffold, are used in other fluorescent probes. These dyes are useful due to their increased fluorescence upon binding to, e.g., nucleic acids (e.g., TOTO, YOYO) [105] or to amyloid-like fibrils (e.g., Thioflavin T [106], Nile red [107]).

## 4. Towards a consistent nomenclature for PIFE

As discussed in sections 2 and 3, there are a growing number of variants of PIFE assays with distinct acronyms: NAIFE [75], SIFI [93,94], and UIFE [74]. Importantly, these PIFE variants use the same underlying photophysical phenomenon in different biophysical assays. By convention, we assume that the PIFE dye undergoes photoisomerisation to a weakly emissive state. In general, therefore, an interaction of the dye with a biomolecule causes a net fluorescence enhancement relative to some minimally hindered isomerization state such as for a freely diffusing dye. Additionally, the *fluorescence enhancement* in the PIFE acronym implies an effect relative to a standard isomerisation rate. In fact, the majority of PIFE works report a *fold increase* in fluorescence intensity, emphasising it is a relative measure. In the absence of an absolute reference, similar photoisomerisation rates responsible for fluorescence enhancement in select assays could effectively lead to quenching (see PIFQ in Section 2.3) in other cases [59].

With these considerations at hand, it would be best to describe the methods in terms of a general photophysical effect, irrespective of the many possible mechanisms that lead to fluorescence modulation. The PIFE acronym, however, is now well-established. For that reason, keeping the acronym would be desirable while still being consistent with the earlier considerations. We, therefore, suggest re-naming the methods described here as **p**hoto**i**somerisation-related **f**luorescence **e**nhancement (PIFE), as was recently proposed [108]. Although this definition leaves the molecular origin that modulates photoisomerisation (e.g., specific interactions, steric obstruction, viscosity) undefined, the different PIFE methods could still be specified in a context-dependent manner.

Since fluorescence enhancement is relative to reference samples that are sometimes not well-defined, one should strive to report PIFE results with absolute fluorescence intensity rather than relative *fold changes*. However, due to the arbitrary dependence of the intensity on the excitation power and other sources irrelevant to PIFE, we recommend using similar dyes that lack the capability to photoisomerise (see sCy3 vs. Cy3B; Figure 6A) as controls for the maximum possible fluorescence enhancement. Alternatively, fluorescence lifetimes [9,109–111], which report on PIFE decoupled from any other potential factors that influence fluorescence intensity, can be used since often a non-PIFE reference may not be available.

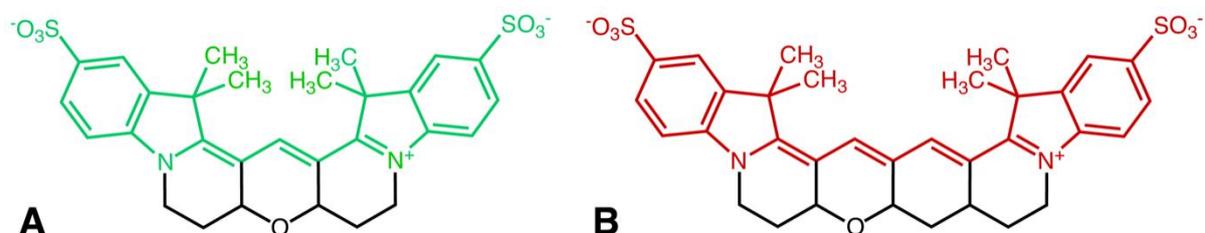

**Figure. 6. Chemical structures of restrained cyanine dyes**. The chemical structures of the rigidified cyanine dyes **(A)** Cy3B and **(B)** Cy5B lack a flexible polymethine chain and do not show any fluorescence-modulating *cis/trans* isomerisation. The chemical structures of the unrestricted sulfo-Cy dyes are highlighted in color.

## 5. PIFE: what's next?

After establishing the initial concept and its application for various biomolecular systems, the remaining question is where to go next? From an applicative perspective, the utilisation of PIFE in various emerging platforms to tackle previously unexplored problems is experiencing a clear surge. These platforms encompass a wide range of innovative techniques, such as, but not limited to, smPIFE with continuous repositioning of single molecules within an anti-Brownian electrokinetic trap [112], the integration of smPIFE with force spectroscopy [113], the application of PIFE for investigating liquid-liquid phase separation of the RNA-binding protein FUsed in Sarcoma (FUS) [114], as well as bio-detection and sensing applications [115,116], including the introduction of aptamer-based PIFE [117,118]. Undoubtedly, this surge in adoptions of smPIFE points towards a promising future of PIFE in the realms of biomolecular science and biomedical application development. In the following section, we discuss important issues that still require attention and also outline some recent developments: (i) avoiding PIFE effects, (ii) modelling and simulating PIFE, (iii) developing new PIFE dyes, (iv) fluorescence lifetime-based single-molecule PIFE burst analysis (v) combining PIFE with FRET, and (vi) using PIFE in cellular imaging.

### 5.1 On avoiding PIFE

Before discussing how PIFE can be exploited in the future, we start by considering situations in which PIFE can or has to be avoided. While the PIFE effect can be a powerful tool, it has the potential to be a confounding and undesired variable in various biomolecular assays. For example, in a FRET experiment, the fluorescence enhancement of a cyanine donor alters the FRET signal, which could be incorrectly interpreted as a distance change. Furthermore, PIFE affects the fluorescence QY of the donor and hence alters the Förster distance, $R_0$, which can result in an incorrect conversion of FRET efficiencies to inter-dye distances.

As such, we should be wary of the PIFE effect when designing non-PIFE fluorescence-based experiments. The guidance here is essentially the opposite of designing a PIFE experiment, that is to avoid using cyanine dyes or, if cyanine dyes are used, to strategically place them such that they will not exhibit a PIFE effect. Conveniently, rigidified bridged cyanine dyes have been and are continuously being developed (e.g., Cy3B, Cy5B; Figure 6) [12,14], which assist in eliminating the photoisomerisation as well as in serving as a control for the maximum PIFE enhancement. For FRET experiments, there are rhodamine-based alternatives to the popular donor Cy3 that have similar spectral and photophysical characteristics but do not exhibit PIFE. For example, the dye ATTO 550 has been shown to work well when conjugated to DNA [119], whereas the dyes Alexa Fluor 546, ATTO 532 and ATTO 643 have been successfully conjugated to proteins for quantitative smFRET studies [120]. Alternatively, one could exploit PIFE to modulate the fluorescence QY of the donor dye. By increasing the QY and hence $R_0$, one could measure longer inter-dye distances using FRET. However, careful determination of the donor's fluorescence QY in the absence of the acceptor would be essential here.

If cyanine dyes are required in non-PIFE fluorescence-based experiments, it is often useful to avoid the PIFE effect by positioning the cyanine dye such that (1) it is not constrained by its environment, and (2) its environment does not change upon the event that is to be observed (i.e., conformational changes, partner binding). As explained below, dye-specific AV calculations can be used to assess the labelling positions of candidate dyes for this purpose. If labelling nucleic acids, it may be preferable to conjugate cyanine dyes to internal bases away from ends to avoid stacking effects onto terminal bases [83,96], or alternatively to use A/T and A/U base-pairs at those ends, and certainly to position these dyes away from protein binding footprints to avoid pPIFE. When labelling a biomolecule that undergoes a conformational change, one should place cyanine dyes away from sites of structural rearrangement. In either case, if one wishes to convert measured FRET efficiencies to absolute inter-dye distances, then it would be required to measure the fluorescence QY [5] of donor cyanine dyes conjugated to the molecule of interest and recalculate $R_0$ using this more accurate value [121].

### 5.2 MD simulations of PIFE

Since PIFE is highly sensitive to the chemical microenvironment of the dye [18,59,93], it is not straightforward to predict the perfect dye-labelling site that will generate a robust PIFE signal. Nevertheless, it is possible to obtain 3D structures, for instance of a nucleic acid-protein complex, through structure determination methods or integrative modelling approaches (compare Figure 7 or 5D-F) [122]. The obtained models serve as the basis for identifying optimal labelling sites for PIFE [100,101]. Different potential labelling sites can be investigated regarding the steric hindrance of the dye in the presence of a protein (Figure 7), which remains the best available predictor of PIFE to date. An approximation of steric hindrance can be determined by the ratio of the AV and the CV of the dyes, which can be obtained with open-source libraries like "LabelLib" [123] or "FRETraj" [89]. In the case of potential dynamic structural ensembles, coarse-grained molecular dynamics (MD) simulations, as well as *de novo* modelling of the biomolecular structure, are applicable for finding equilibrium conformations which, again, serve as the basis for identifying the optimal labelling site for PIFE. Thus, an identified potential labelling site that exhibits a significant change in AV or CV upon protein binding or conformational change can be selected as a good starting candidate for the first PIFE experiment (see Figure 7 or 5D-F).

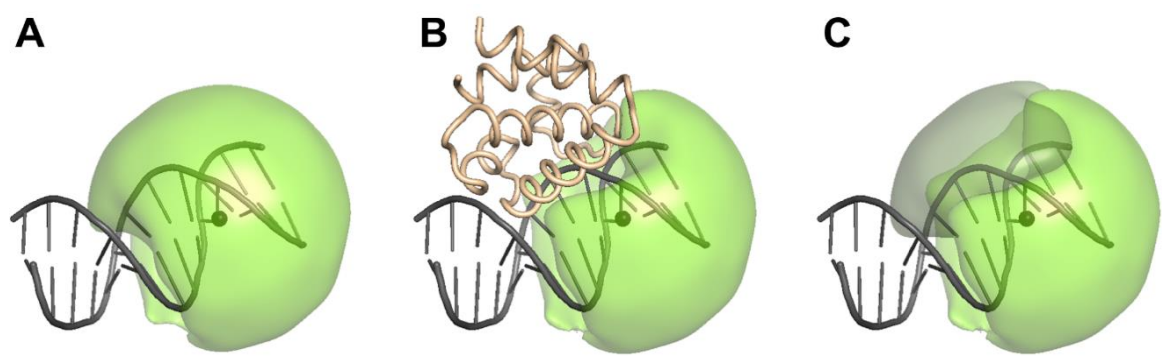

**Figure 7. Choosing the labelling site for the PIFE experiment.** As an example, the binding reaction of a 93-residue bacterial DNA-binding protein domain (brown) to DNA (grey) is shown. The snapshot of the DNA-protein complex is generated by CafeMol, a coarse-grained simulation package[124]. (A) The AV of the dye in a free or "reference" state is displayed by the green surface, while the dye attachment site (dT) on the DNA is indicated by the inner dark sphere. (B) The AV of dye in the bound or "PIFE" state is reduced by the presence of the bound protein, which will likely result in detectable PIFE. (C) The comparison between the AV of the free and bound states (cf. panels A and B) reveals a volume difference of 23% - represented by the grey surface.

However, molecular structures are rarely static. In particular, RNA represents a highly dynamic system [125] that is inadequately represented by a single structure. MD simulations of *in silico* labelled biomolecules have enabled the calculation of photophysical parameters in this context [126–128,89]. With the Python-based FRETraj software package, the calculation of the multi-AV/ACV and the dynamic anisotropy along one or an ensemble of MD trajectories has become possible [89]. This realistically captures the dynamic image of both the dye and the host molecule and allows the comparison of simulated data with experiments (see also Fig. 3). Data generated in this way do not just predict ideal labelling positions but also facilitates the interpretation of PIFE and FRET experiments, as they give an atomistic picture of the structural rearrangements.

In contrast, the simulation of fluorescence lifetime has only been possible through elaborate *ab initio* MD-quantum mechanical (QM) simulations so far and has, therefore, not been considered for modelling PIFE. Also, the photoisomerisation probability of the polymethine chain of Cy dyes is not considered in the current force fields for MD simulations mostly since only the ground-state structure of the dyes, and not those of the excited state, were incorporated into existing MD force fields [129]. Yet, the prediction of the stacking probability of Cy dyes in their local chemical environment is possible by considering the ACV along the MD trajectory. By simulating the dye movement in an *in silico* labelled host molecule, it is possible to visualise the interaction of the dye with its local chemical environment and compare the stacking probability to experimentally accessible parameters such as the dynamic fluorescence anisotropy. This information can be used not only in predicting PIFE but also in correcting the prediction of FRET values when using Cy dyes. Furthermore, combining PIFE and FRET will increase the dynamic range for integrative modelling. PIFE scans the local environment of the dye, weights the ACV, and thereby complements distance constraints obtained by FRET to restrict a *de novo* generated structural ensemble.

### 5.3 New PIFE dyes

As described above, dyes and their properties are at the heart of PIFE-based assays. So far, most PIFE assays are based on cyanine dyes and, in particular, Cy3 and sCy3. Which structural alternatives could serve for future extensions of PIFE assays? In this regard, the specific assay type (e.g., ensemble or single molecule) and other parameters also need to be considered for future optimization in terms of dynamic range, spectral regime, and compatibility with other assays (see PIFE-FRET below). The dynamic range of PIFE is governed by brightness changes between the non-influenced dye (before PIFE occurs) and the fully "restricted" state (with PIFE). These changes can be increased either by lowering the fluorescence QY of the dye or promoting restriction by interaction with the biomolecule. Based on the currently available fluorophores, a fundamental question is by how much these two states should differ. This is especially relevant in light of available base-intercalating or fluorogenic dyes, where fluorescence enhancement factors of up to 1,000-fold can be achieved [130]. In that regard, it is unclear how strong fluorescence suppression in the non-PIFE state should be, knowing that, at least in single-molecule assays, the non-PIFE state (and its photon output) can determine how viable the specific assay is.

In our view, new dyes should thus be developed (or identified) that feature a wide dynamic range for proximity-dependent fluorescence enhancement for use in quantitative PIFE studies. The brightness changes could be gradual or based on many different distinguishable brightness states, or alternatively only switch between the two extreme states (i.e., non-PIFE vs. PIFE) for semi-quantitative assays. The former can be realised by reducing specific interactions between dye and biomolecule, such that the enhancement of fluorescence originates exclusively from the steric restriction. In contrast, the latter 'on-off' PIFE sensor could be achieved by specifically promoting the dye-biomolecule interactions. A target-specific PIFE sensor could be designed by functionalising the core structure of the dye with various side groups to sense different domains based on their charge or hydrophobicity (e.g., through the addition of cholesterol anchors) or even detect specific side chains or post-translational modifications.

Furthermore, the interpretation of PIFE assays, particularly those where quantitative information is desired, could benefit from a clear determination of the maximum PIFE fluorescence enhancement. For this control, dyes that can define the maximum PIFE enhancement, such as rigidified dyes, are required, where no photoisomerisation is possible, as has been described here for Cy3B (see also Figure 6A). This aspect is also relevant for extending PIFE into other spectral regimes, where such desired control dyes are not yet well-established. Here, developments such as the Cy3B [12] and more recently the Cy5B [14] and Cy7B [131] derivatives (Figure 6B) from the Schnerman lab [132] will be important puzzle pieces for designing new PIFE assays. In general, one could also envisage other types of photoswitches that could undergo environment-sensitive changes in fluorescence properties. For this to be achieved, many of the hybrid dyes with photoswitchable properties, such as indigos [133], stilbenes [134], spiropyrans [135] and hemithioindigos [136], might be relevant for tests where the planar isomer often shows larger brightness. Here, fluorescent nucleobase analogous that can photoisomerise, are of particular interest, as they can be easily incorporated in the nucleic acid sequence at the expense of lower fluorescence QYs [137]. An extended palette of dyes with a high dynamic range of the PIFE effects at different spectral regimes may support further combinations of PIFE aspects with other biophysical assays, such as FRET (see discussion below).

### 5.4 Combining PIFE and FRET as a multi-proximity ruler

Single-molecule fluorescence-based assays that can simultaneously read out multiple distances are highly desirable. Such assays can probe correlated conformational changes in multi-domain proteins and complexes, monitor conformational changes at both short and intermediate biomolecular distances, and also visualise binding-induced conformational rearrangement during complex formation. While multi-color FRET approaches can monitor multiple distances simultaneously, they are based on site-specific labelling of at least three fluorescent probes. Labelling with multiple fluorophores, however, is often hampered by (i) low labelling efficiencies, (ii) difficulties in achieving directed site-specific labelling, (iii) too many dye labelling permutations in statistical labelling, (iv) the requirement of a sophisticated FRET analysis, and (v) pure stability and high fluorescence QY of the available dyes, particularly in the UV or NIR ranges. Moreover, such approaches often lack appropriate FRET pairs with short Förster distances to probe short-range distance changes as they occur, for example, during alterations in binding modes.

To address these issues, PIFE-FRET was proposed [138,139] and later realised in immobilised single-molecule assays [139–142], where binding in close proximity to the PIFE-sensitive dye leads to changes in its fluorescence, without subsequent changes to the fluorescence of the acceptor dye, and conformational changes induced anti-correlated changes in donor and acceptor fluorescence due to changes in FRET. In such immobilised single-molecule assays, changes in PIFE are observed in both the donor intensity trajectory and in the total intensity trajectory of the sum of donor and acceptor fluorescence intensity, while changes in FRET are observed through the ratio of the acceptor intensity to total intensity. In parallel to the implementation on immobilised molecules, PIFE-FRET for freely-diffusing molecules was realised using single-molecule microsecond alternating laser excitation (µsALEX) experiments [18,72] for simultaneously monitoring the interaction between nucleic acids and proteins and their associated binding-induced conformational changes. Here, short-range (<3 nm) surface proximity sensing via PIFE for probing the protein-DNA interaction and single-molecule FRET as the readout for any conformational changes in the targeted nucleic acid was introduced. The reporter dye sCy3 was placed at the 5'-end of the dsDNA in proximity (<3 nm) to the binding site of different restriction enzymes (Figure 8A-C). Then, parameters were retrieved from µsALEX to report on the intra-molecular distance between the donor and acceptor dyes, and the inter-molecular proximity of the DNA-binding protein to a Cy dye. The stoichiometry ratio, $S$, was used as a readout for the change in brightness due to PIFE and could confirm the linear distance dependence for the binding-induced fluorescence enhancement after disentangling its contribution to FRET. Importantly, a theoretical framework for the *E-S* dependence in PIFE-FRET experiments was developed [18] and could be employed to report on PIFE and FRET for each subpopulation.

More recently, a proof of principle experiment was presented [143], in which smFRET and PIFE were combined to simultaneously probe conformational changes within single protein domains during their interaction with neighboring protein domains. As an example, the inter- and intra-domain interactions in the tandem substrate-binding domains (SBDs) 1 and 2 of the bacterial ABC import system GlnPQ were visualised ([Figure 8D-F](#)). Here, smFRET served to monitor the conformational state of one domain (SBD2), while PIFE probed its interaction with the neighboring domain SBD1.

While FRET is a directional process occurring from a donor dye to an acceptor dye, PIFE can occur in any of the two dyes. It can occur in the donor dye or in the acceptor dye but also in both dyes at the same time, depending on the type of dyes used. To derive accurate FRET values, and hence inter-dye distances, it is, therefore, desirable to restrict PIFE to one dye only and keep the other dye as an environmentally-insensitive dye. PIFE modulates the photoisomerisation rates and accordingly the fluorescence QY of the affected dye. While acceptor-based PIFE will lead to a shift in the µsALEX stoichiometry ratio, independent of the FRET process that might happen in parallel, donor-based PIFE will lead to an alteration of $R_0$, as well as the $\gamma$ correction factor, which balances the differences in donor and acceptor dye fluorescence QYs and detection efficiencies. Using a single PIFE-sensitive dye, it was shown that it is possible to disentangle FRET from PIFE ([Figure 8G](#)) and determine the protein-specific PIFE effect as long as only one of the dyes is affected [18,72]. Regardless, due to the independence of FRET and PIFE in acceptor-based PIFE-FRET, such an assay is more desirable. However, potential dyes for PIFE that can act as FRET acceptors usually have excitation and emission spectra shifted to the red relative to sCy3, with more conjugated π bonds within the polymethine chain. This, in turn, might influence the efficiency of photoisomerisation. Indeed, such dyes (e.g., Cy5, Alexa Fluor 647) were tested, and it was found that the dynamic range that can be measured in PIFE using them is smaller relative to that of sCy3 [72]. Alternatively, a combination of any blue donor dye (e.g., Alexa Fluor 488 or ATTO 488) that does not exhibit microenvironment-sensitive fluorescence, with Cy3 or sCy3 as an acceptor dye, might serve as the basis for acceptor-based PIFE-FRET applications. Another possibility in which PIFE-FRET can be useful is for diffusion-based analysis of brightness changes in a PIFE-sensitive dye, assuming that the two FRET dyes are separated on one biomolecule at an inter-dye distance larger than the dynamic range of FRET. In this case, it is possible to use µsALEX to determine PIFE or other fluorescence modulation effects [72]. Here, the stoichiometry ratio, *S*, directly reports on the brightness change of one dye using the constant intensity of the acceptor dye as an internal reference.

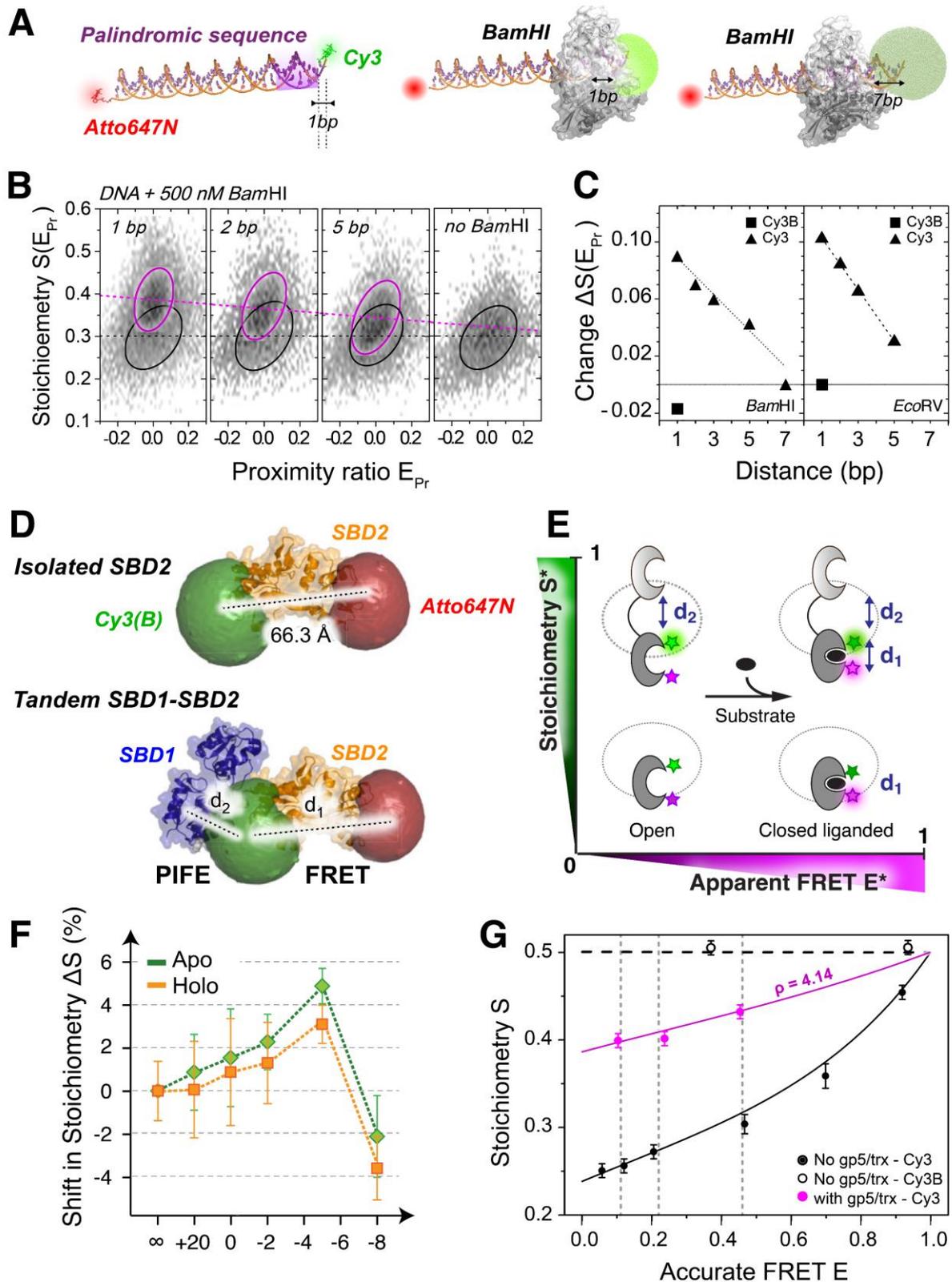

**Figure 8. Single-molecule PIFE-FRET monitored by μsALEX spectroscopy. A-C)** Protein-nucleic acid interaction. **(A)** Dye AV calculation for sCy3 attached at the 5'-end in the presence of BamHI bound to dsDNA. **(B)** E-S 2D histograms and **(C)** Stoichiometry change due to PIFE for BamHI and EcoRV bound to dsDNA as a function of the proximity of the sCy3 dye from the palindromic binding sequence. **D-F)** Protein-protein interaction probed by PIFE-FRET between substrate-binding domains 1 and 2 of the bacterial ABC importer GlnPQ. **(D)** Assay for SBD2 as an isolated domain and in tandem with SBD1. **(E)** Working principles to probe conformations and interaction between SBD1 and SBD2 via PIFE-FRET. **(F)** PIFE occurs between both domains for shortened linker length in the open and substrate-bound state of SBD2. **(G)** Disentangling of PIFE and FRET in PIFE-FRET assays. Accurate FRET and PIFE-enhancement for BamHI and the polymerase gp5/trx and BamHI on dsDNA. Licenses: A-C,) Reprinted from Ploetz, Lerner *et al.* [72] under the terms of an ACS AuthorChoice License. D-F) Reprinted from Ploetz, Schuurman-Wolters *et al.* [143] under the terms of the Creative Commons CC-BY License 4.0. G) Reprinted from Lerner, Ploetz *et al.* [18] under the terms of the Creative Commons CC-BY License 4.0.

Despite the difficulties in PIFE-FRET experiments where both dyes can be PIFE-sensitive, such as the combination of a Cy3 donor and a Cy5 acceptor, recent reports show the usefulness of this approach, at least in immobilised single-molecule assays, in sensing binding through both the donor dye and the acceptor dye, while reporting on conformational changes through FRET [144]. Recently, the quantitative interpretation of such donor-and-acceptor-PIFE-FRET experiments has been challenged through the use of a hidden Markov model approach was suggested for the tandem analysis of both donor and acceptor PIFE changes and FRET changes [145].

**5.5 Lifetime-based PIFE**

In almost all intensity-based assays, PIFE is assessed as a relative effect. This is seen through the requirement to calibrate the values of relative stoichiometry ratios in PIFE-FRET. Alternatively, smPIFE measurements can be performed without relying on fluorescence intensities, analogous to the approach taken previously at the ensemble level using fluorescence lifetimes (see Figures 1 and 3 and section 3.2). As described earlier, lifetime-based PIFE removes the reliance on the intensity parameter, which could be affected by parameters other than photoisomerisation. Under the assumption that the major reason for modulating the mean fluorescence lifetime is due to changes in photoisomerisation, lifetime-based assays can report solely on PIFE effects relative to a minimal mean fluorescence lifetime value. Therefore, even in lifetime-based PIFE, the results are not absolute but relative to some basal fluorescence lifetime values that are most probably system-specific.

For single-molecule experiments, a parameter equivalent to the mean fluorescence lifetime of the PIFE dye, the mean photon nanotime, can be reported per each single-molecule photon burst (Figure 9). Indeed, such lifetime-based smPIFE studies have emerged, in which the reported data is shown as a histogram of sCy3 mean nanotimes per each single-molecule burst [109,110]. These histograms often exhibit sub-populations, which clearly report on instances in which different degrees of PIFE have been reported. Then, upon careful consideration of the results, they can be interpreted as groups of molecules exhibiting different local structures in the vicinity of the sCy3-conjugated residue, leading to different degrees of steric obstruction of the sCy3 excited-state *cis-trans* isomerisation.

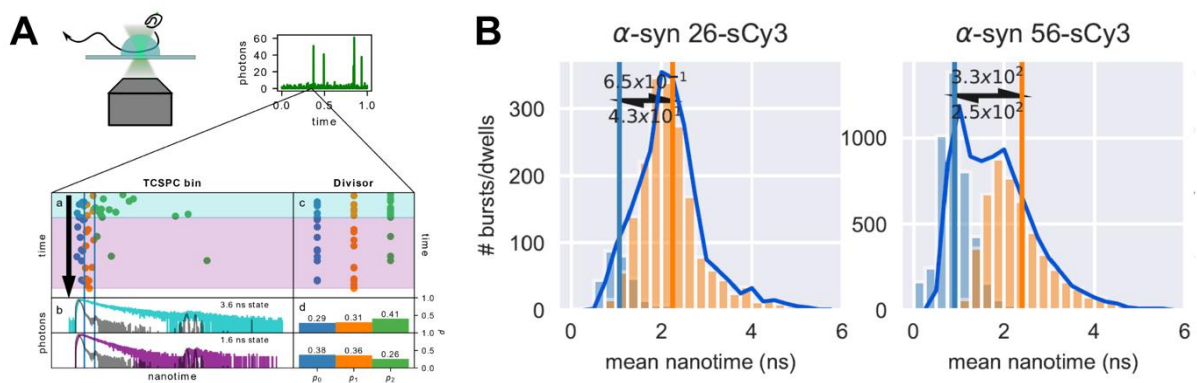

**Figure 9. Probing PIFE within bursts in single-molecule fluorescence experiments. (A)** Divisor-approach for analysing within-burst fluorescence lifetime dynamics. **(B)** MpH$^2$MM analysis of the unbound α-syn monomer labelled at positions 26 and 56 with sCy3 provides histograms of mean nanotimes of state dwells. License: Reprinted from Harris PD & Lerner E, "Identification and quantification of within-burst dynamics in singly labelled single-molecule fluorescence lifetime experiments", 2:100071, Copyright (2022) [146], with permission from Elsevier.

SmPIFE, like other single-molecule fluorescence-based detection methods, can be performed with immobilised or diffusing molecules (Figure 9). Immobilised single-molecule assays for tracking the trajectory of the dye fluorescence lifetime were first introduced for smPIFE measurements to monitor slow dynamics between intermediate states in DNA transcription [147]. As for diffusion-based single-molecule assays, the acquisition of burst data from single diffusing molecules is usually performed using point detectors, which provide ps precision on photon nanotimes and typical time between consecutive photons of a few µs. A dye-labelled biomolecule can undergo a conformational transition while traversing the confocal volume. If these different conformations are associated with changes in the degree of PIFE experienced by the dye-labelled residue, and concomitantly with a difference in fluorescence brightness or lifetime, this will result in single-molecule fluorescence bursts exhibiting within-burst dynamics [148–151,146]. In such scenarios, the mean fluorescence brightness or lifetime might not report on the values representing either of the conformations, but rather on a time average of the conformations. Many tools have been developed in the last two decades for analysing smFRET burst measurements and more specifically for identifying and even quantifying the underlying dynamics. These approaches are summarized in recent reviews of the smFRET field [152,153].

Harris and Lerner have extended an approach to identify and quantify within-burst dynamics [154], originally introduced by Haran and co-workers, photon-by-photon hidden Markov modelling (H$^2$MM) [149]. Using this extended approach, termed multi-parameter H$^2$MM (mpH$^2$MM), and inspired by the work of Antonik and co-workers [155], it became possible to quantitatively analyse within-burst fluorescence dynamics of smFRET but also lifetime-based PIFE dynamics [146]. Using this approach, they have shown the unbound α-synuclein monomer exhibits PIFE dynamics on the timescale of a few ms (Figure 9B), which point towards dynamics occurring in the vicinity of and affecting the Cy3-labelled residues [108–110]. It may be envisaged that smPIFE will undergo similar advancements as smFRET did, pushing the limits of PIFE for reporting dynamics of local effects in the 0-3 nm proximity range.

**5.6 Taking PIFE into the cell**

An exciting question is whether PIFE can actually be taken to the cell. The feasibility of this idea is supported by the successful delivery of dye-labelled DNA and proteins into cells via physical methods (e.g., electroporation, microinjection), e.g., for single-molecule studies [156–159]. Furthermore, in-cell labelling has become possible using bio-orthogonal labelling approaches [160] and various self-labelling protein tags [161–164]. However, the feasibility of in-cell PIFE assays to derive meaningful information is impacted by two major factors: (i) cells feature distinct and quite heterogeneous viscosities in different compartments and strongly differ from dilute buffer conditions, (ii) nonspecific and unwanted interactions of macromolecules with the Cy3-conjugated probe are possible within a cell. Consequently, the desired PIFE effects that monitor, for instance, a binding event of the PIFE-probe to its target, need to be distinguished from viscosity-driven effects in different cell compartments or unwanted "background"-binding of the PIFE-probe to other cellular macromolecules.

To solve these problems, one could envision a ratiometric type of PIFE assay where a second dye is used as an internal photophysical standard (as done in PIFE-FRET for large donor-acceptor separations; Figure 8A). An initial mapping of PIFE-related brightness or lifetime ratios in the cells will indicate whether there are cellular regions with variable Cy3-enhancement effects. Such regions may also be highlighted by fluctuations in the signal when it moves inside the cell. To distinguish between the effects of viscosity and interaction, translational diffusion of the sensor, in contrast to immobilised phases, might be used as additional information. Purely viscosity-driven effects should not lead to probe immobilisation and may or may not be linked to a clear diffusion change. Another readout might be a change in the sensor diffusivity if the target is reasonably large relative to the sensor, which may also exhibit a Cy3-PIFE effect due to target proximity. One can also envision similar experiments with labelled antibodies or aptamers binding to their targets. The clarity of the observables and the interpretation will depend on the timescales of the interactions and the diffusivity of molecular sensors and targets.

While the proposed probe designs could be envisioned in distinguishing between nonspecific viscosity effects and specific interaction effects in *in-cell* PIFE measurements, we are sure that other designs can also be considered for specific biological questions and cellular contexts. In general, we believe that there is considerable scope to bring PIFE into the cell, and we believe that examples of intracellular PIFE should emerge in the near future.

## 6. Conclusion

Assays based on PIFE have found widespread use in biophysical and biochemical research. This review has brought together a community of active contributors who discussed the most recent developments and future avenues of this research direction. The main mechanistic aspects of PIFE were summarised, including how it was conceived and developed. Furthermore, the diverse applications of PIFE in biophysical and biochemical assays were discussed showing the bright future of PIFE as a tool to investigate biomolecular structures, their dynamics and interactions. Our work also led to the proposal to change the acronym PIFE to ***p**hoto**i**somerisation-related **f**luorescence **e**nhancement*, reflecting the underlying photochemical mechanism rather than specific applications. We hope that this work will motivate new researchers to contribute to this prospering field through the design, synthesis, and exploitation of new photo-responsive dyes, and the development of novel assays and quantitative approaches using PIFE.

## 7. Acknowledgements

A.B. acknowledges funding from the European Union's Horizon 2020 research and innovation program under the Marie Skłodowska-Curie Grant agreement no. 101029907. R.B., F.E., and F.D.S. received funding from the European Social Funds as REACT-research group [REACT EU – 100602650. T.C. is currently supported by the Center for Nanoscience (CeNS), the Bundesministerium für Bildung und Forschung (KMU grant „quantumFRET") and the Deutsche Forschungsgemeinschaft (CO879/4-1 & CO879/6-1). H.D.K. acknowledges support from the National Institutes of Health (R01GM112882). M.L. acknowledges support by the National Science Foundation [MCB- 1918716]. T.L. acknowledges support by the NIH R35 GM136632. S.W.M. acknowledges support by BBSRC (BB/T003197/1). E.L. acknowledges support by the Israel Science Foundation (ISF 556/22).

## 8. Author Contributions

E.P., T.C., S.W.M. and E.L. wrote the initial manuscript, coordinated the acquisition of the contributions of all co-authors, and wrote the final version of the review. E.P. prepared the figures for this work. All other co-authors contributed to sections of the manuscript: T.L. and M.L. wrote about the history of PIFE. A.B., M.L., T.C., S.W.M. and E.L. wrote about dye photophysics. F.E., F.D.S. and R.B. wrote about NAIFE and prepared the figure about this subject. S.W.M. wrote about SIFI. A.M. and A.N.K. wrote about UIFE. H.D.K., F.E., F.B. and R.B. wrote about modelling PIFE. T.C. wrote about the design principles for PIFE-based dyes. B.A. and D.S.R. wrote about how to avoid PIFE in certain situations. PIFE. E.P., T.C. and E.L. wrote about the combination of PIFE and FRET. E.L. wrote about fluorescence lifetime-based PIFE. A.N.K. and T.C. wrote about in-cell PIFE.

**REFERENCES**

[1]   Lakowicz, J R *Principles of Fluorescence Spectroscopy* (Springer New York, NY)
[2]   Valeur, B and Berberan-Santos, M N 2012 *Molecular Fluorescence: Principles and Applications* (Wiley)
[3]   Borisov S M and Wolfbeis O S 2008 Optical Biosensors *Chem. Rev.* **108** 423–61
[4]   Kozlov A G and Lohman T M 2002 Stopped-Flow Studies of the Kinetics of Single-Stranded DNA Binding and Wrapping around the Escherichia coli SSB Tetramer *Biochemistry* **41** 6032–44
[5]   Luo G, Wang M, Konigsberg W H and Xie X S 2007 Single-molecule and ensemble fluorescence assays for a functionally important conformational change in T7 DNA polymerase *Proceedings of the National Academy of Sciences* **104** 12610–5
[6]   Myong S, Cui S, Cornish P V, Kirchhofer A, Gack M U, Jung J U, Hopfner K-P, and Ha T 2009 Cytosolic Viral Sensor RIG-I Is a 5'-Triphosphate–Dependent Translocase on Double-Stranded RNA *Science* **323** 1070–4
[7]   Levitus M and Ranjit S 2011 Cyanine dyes in biophysical research: the photophysics of polymethine fluorescent dyes in biomolecular environments *Quarterly Reviews of Biophysics* **44** 123–51
[8]   Fischer C J, Tomko E J, Wu C G and Lohman T M 2012 Fluorescence Methods to Study DNA Translocation and Unwinding Kinetics by Nucleic Acid Motors *Spectroscopic Methods of Analysis: Methods and Protocols* ed W M Bujalowski (Totowa, NJ: Humana Press) pp 85–104
[9]   Hwang H and Myong S 2014 Protein induced fluorescence enhancement (PIFE) for probing protein–nucleic acid interactions *Chemical Society reviews* **43** 1221–9
[10]  Di Paolo R E, Scaffardi L B, Duchowicz R and Bilmes G M 1995 Photoisomerization Dynamics and Spectroscopy of the Polymethine Dye DTCI *J. Phys. Chem.* **99** 13796–9
[11]  Duchowicz R, Scaffardi L and Tocho J O 1990 Relaxation processes of singlet excited state of 3,3′-diethyloxadicarbocyanine iodide (DODCI) photoisomer *Chemical Physics Letters* **170** 497–501
[12]  Cooper M, Ebner A, Briggs M, Burrows M, Gardner N, Richardson R and West R 2004 Cy3B™: Improving the Performance of Cyanine Dyes *Journal of Fluorescence* **14** 145–50
[13]  Sanborn M E, Connolly B K, Gurunathan K and Levitus M 2007 Fluorescence Properties and Photophysics of the Sulfoindocyanine Cy3 Linked Covalently to DNA *The Journal of Physical Chemistry B* **111** 11064–74
[14]  Michie M S, Götz R, Franke C, Bowler M, Kumari N, Magidson V, Levitus M, Loncarek J, Sauer M and Schnermann M J 2017 Cyanine Conformational Restraint in the Far-Red Range *Journal of the American Chemical Society* **139** 12406–9
[15]  Jia K, Wan Y, Xia A, Li S, Gong F and Yang G 2007 Characterization of Photoinduced Isomerization and Intersystem Crossing of the Cyanine Dye Cy3 *J. Phys. Chem. A* **111** 1593–7
[16]  Hart S M, Banal J L, Bathe M and Schlau-Cohen G S 2020 Identification of Nonradiative Decay Pathways in Cy3 *J. Phys. Chem. Lett.* **11** 5000–7
[17]  Huang Z, Ji D, Wang S, Xia A, Koberling F, Patting M and Erdmann R 2006 Spectral Identification of Specific Photophysics of Cy5 by Means of Ensemble and Single Molecule Measurements *J. Phys. Chem. A* **110** 45–50
[18]  Lerner E, Ploetz E, Hohlbein J, Cordes T and Weiss S 2016 A Quantitative Theoretical Framework For Protein-Induced Fluorescence Enhancement-Förster-Type Resonance Energy Transfer (PIFE-FRET) *The Journal of Physical Chemistry B* **120** 6401–10
[19]  Stennett E M S, Ciuba M A, Lin S and Levitus M 2015 Demystifying PIFE: The Photophysics Behind the Protein-Induced Fluorescence Enhancement Phenomenon in Cy3 *The Journal of Physical Chemistry Letters* **6** 1819–23
[20]  Sundstroem V and Gillbro T 1982 Viscosity-dependent isomerization yields of some cyanine dyes. A picosecond laser spectroscopy study *J. Phys. Chem.* **86** 1788–94

[21] Pronkin P and Tatikolov A 2019 Isomerization and Properties of Isomers of Carbocyanine Dyes *Sci* **1**

[22] Kang J, Lhee S, Lee J K, Zare R N and Nam H G 2020 Restricted intramolecular rotation of fluorescent molecular rotors at the periphery of aqueous microdroplets in oil *Scientific Reports* **10** 16859

[23] Widengren J and Schwille P 2000 Characterization of Photoinduced Isomerization and Back-Isomerization of the Cyanine Dye Cy5 by Fluorescence Correlation Spectroscopy *J. Phys. Chem. A* **104** 6416–28

[24] Zechmeister L and Pinckard J H 1953 On stereoisomerism in the cyanine dye series *Experientia* **9** 16–7

[25] Scheibe G, Heiss J and Feldmann K 1966 Zur photochemischen trans-cis-Umlagerung einfacher Cyaninfarbstoffe *Berichte der Bunsengesellschaft für physikalische Chemie* **70** 52–60

[26] Åkesson E, Sundström V and Gillbro T 1985 Solvent-dependent barrier heights of excited-state photoisomerization reactions *Chemical Physics Letters* **121** 513–22

[27] Korppi-Tommola J E I, Hakkarainen A, Hukka T and Subbi J 1991 An isomerization reaction of a cyanine dye in n-alcohols: microscopic friction and an excited-state barrier crossing *J. Phys. Chem.* **95** 8482–91

[28] Sibbett W, Taylor J, and Welford D 1981 Substituent and environmental effects on the picosecond lifetimes of the polymethine cyanine dyes *IEEE Journal of Quantum Electronics* **17** 500–9

[29] Åkesson E, Hakkarainen A, Laitinen E, Helenius V, Gillbro T, Korppi-Tommola J and Sundström V 1991 Analysis of microviscosity and reaction coordinate concepts in isomerization dynamics described by Kramers' theory *J. Chem. Phys.* **95** 6508–23

[30] Aramendia P F, Negri R M and Roman E S 1994 Temperature Dependence of Fluorescence and Photoisomerization in Symmetric Carbocyanines.Influence of Medium Viscosity and Molecular Structure *J. Phys. Chem.* **98** 3165–73

[31] Murphy S, Sauerwein B, Drickamer H G and Schuster G B 1994 Spectroscopy of cyanine dyes in fluid solution at atmospheric and high pressure: the effect of viscosity on nonradiative processes *J. Phys. Chem.* **98** 13476–80

[32] Mishra A, Behera R K, Behera P K, Mishra B K and Behera G B 2000 Cyanines during the 1990s: A Review *Chem. Rev.* **100** 1973–2012

[33] Mujumdar R B, Ernst L A, Mujumdar S R, Lewis C J and Waggoner A S 1993 Cyanine dye labeling reagents: Sulfoindocyanine succinimidyl esters *Bioconjugate Chem.* **4** 105–11

[34] Gebhardt C, Lehmann M, Reif M M, Zacharias M, Gemmecker G and Cordes T 2021 Molecular and Spectroscopic Characterization of Green and Red Cyanine Fluorophores from the Alexa Fluor and AF Series *ChemPhysChem* **22** 1566–83

[35] Kricka L J 2002 Stains, labels and detection strategies for nucleic acids assays *Ann Clin Biochem* **39** 114–29

[36] Yarmoluk S, Kovalska V and Losytskyy M 2008 Symmetric cyanine dyes for detecting nucleic acids *Biotechnic & Histochemistry* **83** 131–45

[37] Wlodkowic D, Telford W, Skommer J and Darzynkiewicz Z 2011 Chapter 4 - Apoptosis and Beyond: Cytometry in Studies of Programmed Cell Death *Methods in Cell Biology* vol 103, ed Z Darzynkiewicz, E Holden, A Orfao, W Telford and D Wlodkowic (Academic Press) pp 55–98

[38] Hwang G T 2018 Single-Labeled Oligonucleotides Showing Fluorescence Changes upon Hybridization with Target Nucleic Acids *Molecules* **23**

[39] Ma X, Shi L, Zhang B, Liu L, Fu Y and Zhang X 2022 Recent advances in bioprobes and biolabels based on cyanine dyes *Analytical and Bioanalytical Chemistry* **414** 4551–73

[40] Kozlov A G, Galletto R and Lohman T M 2012 SSB–DNA Binding Monitored by Fluorescence Intensity and Anisotropy *Single-Stranded DNA Binding Proteins: Methods and Protocols* ed J L Keck (Totowa, NJ: Humana Press) pp 55–83

[41] Janiak F, Dell V A, Abrahamson J K, Watson B S, Miller D L and Johnson A E 1990 Fluorescence characterization of the interaction of various transfer RNA species with

elongation factor Tu.cntdot.GTP: evidence for a new functional role for elongation factor Tu in protein biosynthesis *Biochemistry* **29** 4268–77

[42] Rajendran S, Jezewska M J and Bujalowski W 2001 Multiple-Step Kinetic Mechanisms of the ssDNA Recognition Process by Human Polymerase β in Its Different ssDNA Binding Modes *Biochemistry* **40** 11794–810

[43] Fischer C J, Maluf N K and Lohman T M 2004 Mechanism of ATP-dependent Translocation of E.coli UvrD Monomers Along Single-stranded DNA *Journal of Molecular Biology* **344** 1287–309

[44] Tomko E J, Fischer C J, Niedziela-Majka A and Lohman T M 2007 A Nonuniform Stepping Mechanism for E. coli UvrD Monomer Translocation along Single-Stranded DNA *Molecular Cell* **26** 335–47

[45] Brendza Katherine M., Cheng Wei, Fischer Christopher J., Chesnik Marla A., Niedziela-Majka Anita, and Lohman Timothy M. 2005 Autoinhibition of Escherichia coli Rep monomer helicase activity by its 2B subdomain *Proceedings of the National Academy of Sciences* **102** 10076–81

[46] Wu C G, Bradford C and Lohman T M 2010 Escherichia coli RecBC helicase has two translocase activities controlled by a single ATPase motor *Nature Structural & Molecular Biology* **17** 1210–7

[47] Wu C G, Xie F and Lohman T M 2012 The Primary and Secondary Translocase Activities within E. coli RecBC Helicase Are Tightly Coupled to ATP Hydrolysis by the RecB Motor *Journal of Molecular Biology* **423** 303–14

[48] Xie F, Wu C G, Weiland E and Lohman T M 2013 Asymmetric Regulation of Bipolar Single-stranded DNA Translocation by the Two Motors within Escherichia coli RecBCD Helicase *Journal of Biological Chemistry* **288** 1055–64

[49] Niedziela-Majka A, Chesnik M A, Tomko E J and Lohman T M 2007 Bacillus stearothermophilus PcrA Monomer Is a Single-stranded DNA Translocase but Not a Processive Helicase in Vitro *Journal of Biological Chemistry* **282** 27076–85

[50] Antony E, Tomko E J, Xiao Q, Krejci L, Lohman T M and Ellenberger T 2009 Srs2 Disassembles Rad51 Filaments by a Protein-Protein Interaction Triggering ATP Turnover and Dissociation of Rad51 from DNA *Molecular Cell* **35** 105–15

[51] Galletto R and Tomko E J 2013 Translocation of Saccharomyces cerevisiae Pif1 helicase monomers on single-stranded DNA *Nucleic Acids Research* **41** 4613–27

[52] Lucius A L, Wong C J and Lohman T M 2004 Fluorescence Stopped-flow Studies of Single Turnover Kinetics of E.coli RecBCD Helicase-catalyzed DNA Unwinding *Journal of Molecular Biology* **339** 731–50

[53] Hwang H, Kim H and Myong S 2011 Protein induced fluorescence enhancement as a single molecule assay with short distance sensitivity *Proceedings of the National Academy of Sciences* **108** 7414 LP – 7418

[54] Nguyen B, Ciuba M A, Kozlov A G, Levitus M and Lohman T M 2019 Protein Environment and DNA Orientation Affect Protein-Induced Cy3 Fluorescence Enhancement *Biophysical Journal* **117** 66–73

[55] Nguyen B, Sokoloski J, Galletto R, Elson E L, Wold M S and Lohman T M 2014 Diffusion of Human Replication Protein A along Single-Stranded DNA *Journal of Molecular Biology* **426** 3246–61

[56] Markiewicz R P, Vrtis K B, Rueda D and Romano L J 2012 Single-molecule microscopy reveals new insights into nucleotide selection by DNA polymerase I *Nucleic Acids Research* **40** 7975–84

[57] Vrtis K B, Markiewicz R P, Romano L J and Rueda D 2013 Carcinogenic adducts induce distinct DNA polymerase binding orientations *Nucleic Acids Research* **41** 7843–53

[58] Sokoloski Joshua E., Kozlov Alexander G., Galletto Roberto, and Lohman Timothy M. 2016 Chemo-mechanical pushing of proteins along single-stranded DNA *Proceedings of the National Academy of Sciences* **113** 6194–9

[59] Rashid F, Raducanu V-S, Zaher M S, Tehseen M, Habuchi S and Hamdan S M 2019 Initial state of DNA-Dye complex sets the stage for protein induced fluorescence modulation *Nature Communications* **10** 2104–2104

[60] Anderson B J, Larkin C, Guja K and Schildbach J F 2008 Chapter 12 Using Fluorophore-Labeled Oligonucleotides to Measure Affinities of Protein–DNA Interactions *Methods in Enzymology* vol 450 (Academic Press) pp 253–72
[61] Song D, Graham T G W and Loparo J J 2016 A general approach to visualize protein binding and DNA conformation without protein labelling *Nature Communications* **7** 10976–10976
[62] Lass-Napiorkowska A and Heyduk T 2016 Real-Time Observation of Backtracking by Bacterial RNA Polymerase *Biochemistry* **55** 647–58
[63] Heyduk E and Heyduk T 2018 DNA template sequence control of bacterial RNA polymerase escape from the promoter *Nucleic Acids Research* **46** 4469–86
[64] Sreenivasan R, Shkel I A, Chhabra M, Drennan A, Heitkamp S, Wang H-C, Sridevi M A, Plaskon D, McNerney C, Callies K, Cimperman C K and Record M T Jr 2020 Fluorescence-Detected Conformational Changes in Duplex DNA in Open Complex Formation by Escherichia coli RNA Polymerase: Upstream Wrapping and Downstream Bending Precede Clamp Opening and Insertion of the Downstream Duplex *Biochemistry* **59** 1565–81
[65] Singh R K, Jonely M, Leslie E, Rejali N A, Noriega R and Bass B L 2021 Transient kinetic studies of the antiviral Drosophila Dicer-2 reveal roles of ATP in self–nonself discrimination ed J M Berger and M A Marletta *eLife* **10** e65810
[66] Chapman J H, Craig J M, Wang C D, Gundlach J H, Neuman K C and Hogg J R 2022 UPF1 mutants with intact ATPase but deficient helicase activities promote efficient nonsense-mediated mRNA decay *Nucleic Acids Research* **50** 11876–94
[67] Harvey B J and Levitus M 2008 Nucleobase-Specific Enhancement of Cy3 Fluorescence *Journal of Fluorescence* **19** 443
[68] Harvey B J, Perez C and Levitus M 2009 DNA sequence-dependent enhancement of Cy3 fluorescence *Photochem. Photobiol. Sci.* **8** 1105–10
[69] Marcia Levitus 2010 Sequence-dependent photophysical properties of Cy3-labeled DNA Proc.SPIE vol 7576
[70] Qiu Y and Myong S 2016 Chapter Two - Single-Molecule Imaging With One Color Fluorescence *Single-Molecule Enzymology: Fluorescence-Based and High-Throughput Methods* vol 581, ed M Spies and Y R B T-M in E Chemla (Academic Press) pp 33–51
[71] Valuchova S, Fulnecek J, Petrov A P, Tripsianes K and Riha K 2016 A rapid method for detecting protein-nucleic acid interactions by protein induced fluorescence enhancement *Scientific Reports* **6** 39653
[72] Ploetz E, Lerner E, Husada F, Roelfs M, Chung S, Hohlbein J, Weiss S and Cordes T 2016 Förster Resonance Energy Transfer and Protein-Induced Fluorescence Enhancement as Synergetic Multi-Scale Molecular Rulers *Scientific Reports* **6** 33257–33257
[73] Kim H, Lee C Y, Song J, Yoon J, Park K S and Park H G 2018 Protein-induced fluorescence enhancement for a simple and universal detection of protein/small molecule interactions *RSC Advances* **8** 39913–7
[74] Mazumder A, Ebright R H and Kapanidis A N 2021 Transcription initiation at a consensus bacterial promoter proceeds via a 'bind-unwind-load-and-lock' mechanism ed M Spies, C Wolberger and N J Savery *eLife* **10** e70090–e70090
[75] Steffen F D, Sigel R K O and Börner R 2016 An atomistic view on carbocyanine photophysics in the realm of RNA *Physical Chemistry Chemical Physics* **18** 29045–55
[76] Zhao M, Steffen F D, Börner R, Schaffer M F, Sigel R K O and Freisinger E 2017 Site-specific dual-color labeling of long RNAs for single-molecule spectroscopy *Nucleic Acids Research* **46** e13–e13
[77] Steffen F D ; B Richard; Freisinger, Eva; Sigel, Roland K O 2019 Stick, Flick, Click: DNA-guided Fluorescent Labeling of Long RNA for Single-molecule FRET *CHIMIA International Journal for Chemistry* **73**
[78] Hanspach G, Trucks S and Hengesbach M 2019 Strategic labelling approaches for RNA single-molecule spectroscopy *RNA Biology* **16** 1119–32

[79]  Cooper D, Uhm H, Tauzin L J, Poddar N and Landes C F 2013 Photobleaching Lifetimes of Cyanine Fluorophores Used for Single-Molecule Förster Resonance Energy Transfer in the Presence of Various Photoprotection Systems *ChemBioChem* **14** 1075–80

[80]  Lee W, von Hippel P H and Marcus A H 2014 Internally labeled Cy3/Cy5 DNA constructs show greatly enhanced photo-stability in single-molecule FRET experiments *Nucleic Acids Research* **42** 5967–77

[81]  Hall L M, Gerowska M and Brown T 2012 A highly fluorescent DNA toolkit: synthesis and properties of oligonucleotides containing new Cy3, Cy5 and Cy3B monomers *Nucleic Acids Research* **40** e108–e108

[82]  Kretschy N and Somoza M M 2014 Comparison of the Sequence-Dependent Fluorescence of the Cyanine Dyes Cy3, Cy5, DyLight DY547 and DyLight DY647 on Single-Stranded DNA *PLOS ONE* **9** e85605

[83]  Spiriti J, Binder J K, Levitus M and van der Vaart A 2011 Cy3-DNA Stacking Interactions Strongly Depend on the Identity of the Terminal Basepair *Biophysical Journal* **100** 1049–57

[84]  Urnavicius L, McPhee S A, Lilley D M J and Norman D G 2012 The Structure of Sulfoindocarbocyanine 3 Terminally Attached to dsDNA via a Long, Flexible Tether *Biophysical Journal* **102** 561–8

[85]  Freisinger E and Sigel R K O 2007 From nucleotides to ribozymes—A comparison of their metal ion binding properties *Coordination Chemistry Reviews* **251** 1834–51

[86]  Sobhy M A, Tehseen M, Takahashi M, Bralić A, De Biasio A and Hamdan S M 2021 Implementing fluorescence enhancement, quenching, and FRET for investigating flap endonuclease 1 enzymatic reaction at the single-molecule level *Computational and Structural Biotechnology Journal* **19** 4456–71

[87]  Lee J M, Kim C R, Kim S, Min J, Lee M-H and Lee S 2021 Mix-and-read, one-minute SARS-CoV-2 diagnostic assay: development of PIFE-based aptasensor *Chem. Commun.* **57** 10222–5

[88]  Steffen F D, Khier M, Kowerko D, Cunha R A, Börner R and Sigel R K O 2020 Metal ions and sugar puckering balance single-molecule kinetic heterogeneity in RNA and DNA tertiary contacts *Nature Communications* **11** 104–104

[89]  Steffen F D, Sigel R K O and Börner R 2021 FRETraj: integrating single-molecule spectroscopy with molecular dynamics *Bioinformatics* **37** 3953–5

[90]  Lakowicz J R 2006 Time-Dependent Anisotropy Decays *Principles of Fluorescence Spectroscopy* (Boston, MA: Springer US) pp 383–412

[91]  Muschielok A, Andrecka J, Jawhari A, Brückner F, Cramer P and Michaelis J 2008 A nano-positioning system for macromolecular structural analysis *Nature Methods* **5** 965–71

[92]  Kalinin S, Peulen T, Sindbert S, Rothwell P J, Berger S, Restle T, Goody R S, Gohlke H and Seidel C A M 2012 A Toolkit and Benchmark Study for FRET-Restrained High-Precision Structural Modeling *Nature Methods* **9** 1218–25

[93]  Morten M J, Lopez S G, Steinmark I E, Rafferty A and Magennis S W 2018 Stacking-induced fluorescence increase reveals allosteric interactions through DNA *Nucleic Acids Research* **46** 11618–26

[94]  Morten M J, Steinmark I E and Magennis S W 2020 Probing DNA Dynamics: Stacking-Induced Fluorescence Increase (SIFI) versus FRET *ChemPhotoChem* **4** 664–7

[95]  Norman D G, Grainger R J, Uhrín D and Lilley D M J 2000 Location of Cyanine-3 on Double-Stranded DNA: Importance for Fluorescence Resonance Energy Transfer Studies *Biochemistry* **39** 6317–24

[96]  Iqbal A, Arslan S, Okumus B, Wilson T J, Giraud G, Norman D G, Ha T and Lilley D M J 2008 Orientation dependence in fluorescent energy transfer between Cy3 and Cy5 terminally attached to double-stranded nucleic acids *Proceedings of the National Academy of Sciences* **105** 11176 LP – 11181

[97]  Søndergaard S, Aznauryan M, Haustrup E K, Schiøtt B, Birkedal V and Corry B 2015 Dynamics of Fluorescent Dyes Attached to G-Quadruplex DNA and their Effect on FRET Experiments *ChemPhysChem* **16** 2562–70

[98] Kim S, Broströmer E, Xing D, Jin J, Chong S, Ge H, Wang S, Gu C, Yang L, Gao Y Q, Su X, Sun Y and Xie X S 2013 Probing Allostery Through DNA *Science* **339** 816 LP – 819

[99] Rosenblum G, Elad N, Rozenberg H, Wiggers F, Jungwirth J and Hofmann H 2021 Allostery through DNA drives phenotype switching *Nature Communications* **12** 2967

[100] Ko J and Heyduk T 2014 Kinetics of promoter escape by bacterial RNA polymerase: effects of promoter contacts and transcription bubble collapse *Biochemical Journal* **463** 135–44

[101] Feklistov A, Bae B, Hauver J, Lass-Napiorkowska A, Kalesse M, Glaus F, Altmann K-H, Heyduk T, Landick R and Darst S A 2017 RNA polymerase motions during promoter melting *Science* **356** 863–6

[102] Fang C, Li L, Shen L, Shi J, Wang S, Feng Y and Zhang Y 2019 Structures and mechanism of transcription initiation by bacterial ECF factors *Nucleic Acids Research* **47** 7094–104

[103] He D, You L, Wu X, Shi J, Wen A, Yan Z, Mu W, Fang C, Feng Y and Zhang Y 2022 Pseudomonas aeruginosa SutA wedges RNAP lobe domain open to facilitate promoter DNA unwinding *Nature Communications* **13** 4204

[104] Koh H R, Roy R, Sorokina M, Tang G-Q, Nandakumar D, Patel S S and Ha T 2018 Correlating Transcription Initiation and Conformational Changes by a Single-Subunit RNA Polymerase with Near Base-Pair Resolution *Molecular Cell* **70** 695-706.e5

[105] Hirons G T, Fawcett J J and Crissman H A 1994 TOTO and YOYO: New very bright fluorochromes for DNA content analyses by flow cytometry *Cytometry* **15** 129–40

[106] Xue C, Lin T Y, Chang D and Guo Z 2017 Thioflavin T as an amyloid dye: fibril quantification, optimal concentration and effect on aggregation *Royal Society Open Science* **4** 160696

[107] Nedaei H, Saboury A A, Zolmajd Haghighi Z and Ghasemi A 2018 Nile red compensates for thioflavin T assay biased in the presence of curcumin *Journal of Luminescence* **195** 1–7

[108] Zamel J, Chen J, Zaer S, Harris P D, Drori P, Lebendiker M, Kalisman N, Dokholyan N V and Lerner E 2021 Structural and Dynamic Insights Into α-Synuclein Dimer Conformations *bioRxiv* 795997

[109] Chen J, Zaer S, Drori P, Zamel J, Joron K, Kalisman N, Lerner E and Dokholyan N V 2021 The structural heterogeneity of α-synuclein is governed by several distinct subpopulations with interconversion times slower than milliseconds *Structure*

[110] Zaer S and Lerner E 2021 Utilizing Time-Resolved Protein-Induced Fluorescence Enhancement to Identify Stable Local Conformations One α-Synuclein Monomer at a Time *JoVE* e62655

[111] Chen J, Zaer S, Drori P, Zamel J, Joron K, Kalisman N, Lerner E and Dokholyan N V 2020 The structural heterogeneity of α-synuclein is governed by several distinct subpopulations with interconversion times slower than milliseconds *bioRxiv* 2020.11.09.374991-2020.11.09.374991

[112] Wilson H, Lee M and Wang Q 2021 Probing DNA-protein interactions using single-molecule diffusivity contrast *Biophysical Reports* **1** 100009

[113] Kacey Mersch, Joshua E. Sokoloski, Binh Nguyen, Roberto Galletto, and Timothy M. Lohman 2022 "Helicase" Activity Promoted Through Dynamic Interactions Between a ssDNA Translocase and a Diffusing SSB Protein *bioRxiv* 2022.09.30.510372

[114] Niaki A G, Sarkar J, Cai X, Rhine K, Vidaurre V, Guy B, Hurst M, Lee J C, Koh H R, Guo L, Fare C M, Shorter J and Myong S 2020 Loss of Dynamic RNA Interaction and Aberrant Phase Separation Induced by Two Distinct Types of ALS/FTD-Linked FUS Mutations *Molecular Cell* **77** 82-94.e4

[115] Mahzabeen F, Vermesh O, Levi J, Tan M, Alam I S, Chan C T, Gambhir S S and Harris J S 2021 Real-time point-of-care total protein measurement with a miniaturized optoelectronic biosensor and fast fluorescence-based assay *Biosensors and Bioelectronics* **180** 112823

[116] Nielsen L D F, Hansen-Bruhn M, Nijenhuis M A D and Gothelf K V 2022 Protein-Induced Fluorescence Enhancement and Quenching in a Homogeneous DNA-Based Assay for Rapid Detection of Small-Molecule Drugs in Human Plasma *ACS Sens.* **7** 856–65

[117] Umrao S, Jain V, Anusha, Chakraborty B and Roy R 2018 Protein-induced fluorescence enhancement as aptamer sensing mechanism for thrombin detection *Sensors and Actuators B: Chemical* **267** 294–301

[118] Song J, Kim H, Lee C Y, Yoon J, Yoo W S and Park H G 2021 Identification of thyroid hormone/thyroid hormone receptor interaction based on aptamer-assisted protein-induced fluorescence enhancement *Biosensors and Bioelectronics* **191** 113444

[119] Hellenkamp B, Schmid S, Doroshenko O, Opanasyuk O, Kühnemuth R, Rezaei Adariani S, Ambrose B, Aznauryan M, Barth A, Birkedal V, Bowen M E, Chen H, Cordes T, Eilert T, Fijen C, Gebhardt C, Götz M, Gouridis G, Gratton E, Ha T, Hao P, Hanke C A, Hartmann A, Hendrix J, Hildebrandt L L, Hirschfeld V, Hohlbein J, Hua B, Hübner C G, Kallis E, Kapanidis A N, Kim J-Y, Krainer G, Lamb D C, Lee N K, Lemke E A, Levesque B, Levitus M, McCann J J, Naredi-Rainer N, Nettels D, Ngo T, Qiu R, Robb N C, Röcker C, Sanabria H, Schlierf M, Schröder T, Schuler B, Seidel H, Streit L, Thurn J, Tinnefeld P, Tyagi S, Vandenberk N, Vera A M, Weninger K R, Wünsch B, Yanez-Orozco I S, Michaelis J, Seidel C A M, Craggs T D and Hugel T 2018 Precision and accuracy of single-molecule FRET measurements—a multi-laboratory benchmark study *Nature Methods* **15** 669–76

[120] Agam G, Gebhardt C, Popara M, Mächtel R, Folz J, Ambrose B, Chamachi N, Chung S Y, Craggs T D, de Boer M, Grohmann D, Ha T, Hartmann A, Hendrix J, Hirschfeld V, Hübner C G, Hugel T, Kammerer D, Kang H-S, Kapanidis A N, Krainer G, Kramm K, Lemke E, Lerner E, Margeat E, Martens K, Michaelis J, Mitra J, Moya Muñoz G G, Quast R, Robb N B, Sattler M, Schlierf M, Schneider J, Schröder T, Sefer A, Tan P S, Thurn J, Tinnefeld P, van Noort J, Weiss S, Wendler N, Zijlstra N, Barth A, Seidel C A M, Lamb D C and Cordes T 2022 Reliability and accuracy of single-molecule FRET studies for characterization of structural dynamics and distances in proteins *bioRxiv* 2022.08.03.502619

[121] Rasnik I, McKinney S A and Ha T 2006 Nonblinking and long-lasting single-molecule fluorescence imaging *Nature Methods* **3** 891–3

[122] Vallat B, Webb B, Fayazi M, Voinea S, Tangmunarunkit H, Ganesan S J, Lawson C L, Westbrook J D, Kesselman C, Sali A and Berman H M 2021 New system for archiving integrative structures *Acta Crystallographica Section D* **77** 1486–96

[123] Dimura M, Peulen T O, Sanabria H, Rodnin D, Hemmen K, Hanke C A, Seidel C A M and Gohlke H 2020 Automated and optimally FRET-assisted structural modeling *Nature Communications* **11** 1–14

[124] Kenzaki H, Koga N, Hori N, Kanada R, Li W, Okazaki K, Yao X-Q and Takada S 2011 CafeMol: A Coarse-Grained Biomolecular Simulator for Simulating Proteins at Work *J. Chem. Theory Comput.* **7** 1979–89

[125] Börner R, Kowerko D, Miserachs H G, Schaffer M F and Sigel R K O 2016 Metal ion induced heterogeneity in RNA folding studied by smFRET *Coordination Chemistry Reviews* **327–328** 123–42

[126] Schröder G F, Alexiev U and Grubmüller H 2005 Simulation of Fluorescence Anisotropy Experiments: Probing Protein Dynamics *Biophysical Journal* **89** 3757–70

[127] Hoefling M, Lima N, Haenni D, Seidel C A M, Schuler B and Grubmüller H 2011 Structural Heterogeneity and Quantitative FRET Efficiency Distributions of Polyprolines through a Hybrid Atomistic Simulation and Monte Carlo Approach *PLOS ONE* **6** e19791–e19791

[128] Hoefling M and Grubmüller H 2013 In Silico FRET from Simulated Dye Dynamics *Computer Physics Communications* **184** 841–52

[129] Graen T, Hoefling M and Grubmüller H 2014 AMBER-DYES: Characterization of charge fluctuations and force field parameterization of fluorescent dyes for molecular dynamics simulations *Journal of Chemical Theory and Computation* **10** 5505–12

[130] Dziuba D, Didier P, Ciaco S, Barth A, Seidel C A M and Mély Y 2021 Fundamental photophysics of isomorphic and expanded fluorescent nucleoside analogues *Chem. Soc. Rev.* **50** 7062–107
[131] Matikonda S S, Hammersley G, Kumari N, Grabenhorst L, Glembockyte V, Tinnefeld P, Ivanic J, Levitus M and Schnermann M J 2020 Impact of Cyanine Conformational Restraint in the Near-Infrared Range *Journal of Organic Chemistry* **85** 5907–15
[132] Eiring P, McLaughlin R, Matikonda S S, Han Z, Grabenhorst L, Helmerich D A, Meub M, Beliu G, Luciano M, Bandi V, Zijlstra N, Shi Z-D, Tarasov S G, Swenson R, Tinnefeld P, Glembockyte V, Cordes T, Sauer M and Schnermann M J 2021 Targetable Conformationally Restricted Cyanines Enable Photon-Count-Limited Applications** *Angewandte Chemie International Edition* **60** 26685–93
[133] Petermayer C and Dube H 2018 Indigoid Photoswitches: Visible Light Responsive Molecular Tools *Acc. Chem. Res.* **51** 1153–63
[134] Waldeck D H 1991 Photoisomerization dynamics of stilbenes *Chem. Rev.* **91** 415–36
[135] Irie M, Fukaminato T, Matsuda K and Kobatake S 2014 Photochromism of Diarylethene Molecules and Crystals: Memories, Switches, and Actuators *Chem. Rev.* **114** 12174–277
[136] Kitzig S, Thilemann M, Cordes T and Rück-Braun K 2016 Light-Switchable Peptides with a Hemithioindigo Unit: Peptide Design, Photochromism, and Optical Spectroscopy *ChemPhysChem* **17** 1252–63
[137] Karimi A, Börner R, Mata G and Luedtke N W 2020 A Highly Fluorescent Nucleobase Molecular Rotor *J. Am. Chem. Soc.* **142** 14422–6
[138] Hohlbein J, Craggs T D and Cordes T 2014 Alternating-laser excitation: single-molecule FRET and beyond *Chemical Society Reviews* **43** 1156–71
[139] Gidi Y, Götte M and Cosa G 2017 Conformational Changes Spanning Angstroms to Nanometers via a Combined Protein-Induced Fluorescence Enhancement–Förster Resonance Energy Transfer Method *J. Phys. Chem. B* **121** 2039–48
[140] Koh H R, Xing L, Kleiman L and Myong S 2014 Repetitive RNA unwinding by RNA helicase A facilitates RNA annealing *Nucleic Acids Research* **42** 8556–64
[141] Koh H R, Kidwell M A, Doudna J and Myong S 2017 RNA Scanning of a Molecular Machine with a Built-in Ruler *J. Am. Chem. Soc.* **139** 262–8
[142] Rodgers M L, O'Brien B and Woodson S A 2023 Small RNAs and Hfq capture unfolded RNA target sites during transcription *Molecular Cell* **83** 1489-1501.e5
[143] Ploetz E, Schuurman-Wolters G K, Zijlstra N, Jager A W, Griffith D A, Guskov A, Gouridis G, Poolman B and Cordes T Structural and biophysical characterization of the tandem substrate-binding domains of the ABC importer GlnPQ *Open Biology* **11** 200406
[144] Song E, Hwang S, Munasingha P R, Seo Y-S, Kang J Y, Kang C and Hohng S 2023 Transcriptional pause extension benefits the stand-by rather than catch-up Rho-dependent termination *Nucleic Acids Research* **51** 2778–89
[145] Sina Jazani and Taekjip Ha 2023 Fluorescence lifetime analysis of smFRET with contribution of PIFE on donor and acceptor *bioRxiv* 2023.04.03.535482
[146] Harris P D and Lerner E 2022 Identification and quantification of within-burst dynamics in singly labeled single-molecule fluorescence lifetime experiments *Biophysical Reports* **2** 100071
[147] Sorokina M, Koh H-R, Patel S S and Ha T 2009 Fluorescent Lifetime Trajectories of a Single Fluorophore Reveal Reaction Intermediates During Transcription Initiation *Journal of the American Chemical Society* **131** 9630–1
[148] Torella J P, Holden S J, Santoso Y, Hohlbein J and Kapanidis A N 2011 Identifying Molecular Dynamics in Single-Molecule FRET Experiments with Burst Variance Analysis. *Biophysical journal* **100** 1568–77
[149] Pirchi M, Tsukanov R, Khamis R, Tomov T E, Berger Y, Khara D C, Volkov H, Haran G and Nir E 2016 Photon-by-Photon Hidden Markov Model Analysis for Microsecond Single-Molecule FRET Kinetics *The Journal of Physical Chemistry B* **120** 13065–75

[150] Schrimpf W, Barth A, Hendrix J and Lamb D C 2018 PAM: A Framework for Integrated Analysis of Imaging, Single-Molecule, and Ensemble Fluorescence Data *Biophysical Journal* **114** 1518–28

[151] Barth A, Opanasyuk O, Peulen T-O, Felekyan S, Kalinin S, Sanabria H and Seidel C A M 2022 Unraveling multi-state molecular dynamics in single-molecule FRET experiments. I. Theory of FRET-lines *The Journal of Chemical Physics* **156** 141501–141501

[152] Lerner E, Cordes T, Ingargiola A, Alhadid Y, Chung S, Michalet X and Weiss S 2018 Toward dynamic structural biology: Two decades of single-molecule Förster resonance energy transfer *Science* **359** eaan1133–eaan1133

[153] Lerner E, Barth A, Hendrix J, Ambrose B, Birkedal V, Blanchard S C, Börner R, Sung Jung H, Cordes T, Craggs T D, Deniz A A, Diao J, Fei J, Gonzalez R L, Gopich I V, Ha T, Hanke C A, Haran G, Hatzakis N S, Hohng S, Hong S-C, Hugel T, Ingargiola A, Joo C, Kapanidis A N, Kim H D, Laurence T, Lee N K, Lee T-H, Lemke E A, Margeat E, Michaelis J, Michalet X, Myong S, Nettels D, Peulen T-O, Ploetz E, Razvag Y, Robb N C, Schuler B, Soleimaninejad H, Tang C, Vafabakhsh R, Lamb D C, Seidel C A M and Weiss S 2021 FRET-based dynamic structural biology: Challenges, perspectives and an appeal for open-science practices ed O Boudker *eLife* **10** e60416–e60416

[154] Harris P D, Narducci A, Gebhardt C, Cordes T, Weiss S and Lerner E 2022 Multi-parameter photon-by-photon hidden Markov modeling *Nature Communications* **13** 1000–1000

[155] Kim G-H, Legresley S E, Snyder N, Aubry P D and Antonik M 2011 Single-Molecule Analysis and Lifetime Estimates of Heterogeneous Low-Count-Rate Time-Correlated Fluorescence Data *Applied Spectroscopy* **65** 981–90

[156] Sakon J J and Weninger K R 2010 Detecting the conformation of individual proteins in live cells *Nature Methods* **7** 203–5

[157] Crawford R, Torella J P, Aigrain L, Plochowietz A, Gryte K, Uphoff S and Kapanidis A N 2013 Long-Lived Intracellular Single-Molecule Fluorescence Using Electroporated Molecules *Biophysical Journal* **105** 2439–50

[158] Plochowietz A, Crawford R and Kapanidis A N 2014 Characterization of organic fluorophores for in vivo FRET studies based on electroporated molecules *Physical Chemistry Chemical Physics* **16** 12688–94

[159] König I, Zarrine-Afsar A, Aznauryan M, Soranno A, Wunderlich B, Dingfelder F, Stüber J C, Plückthun A, Nettels D and Schuler B 2015 Single-molecule spectroscopy of protein conformational dynamics in live eukaryotic cells *Nature Methods* **12** 773–9

[160] Yu M, Heidari M, Mikhaleva S, Tan P S, Mingu S, Ruan H, Reinkermeier C D, Obarska-Kosinska A, Siggel M, Beck M, Hummer G and Lemke E A 2022 Deciphering the conformations and dynamics of FG-nucleoporins <em>in situ</em> *bioRxiv* 2022.07.07.499201

[161] Keppler A, Gendreizig S, Gronemeyer T, Pick H, Vogel H and Johnsson K 2003 A general method for the covalent labeling of fusion proteins with small molecules in vivo *Nature Biotechnology* **21** 86–9

[162] Gautier A, Juillerat A, Heinis C, Corrêa I R, Kindermann M, Beaufils F and Johnsson K 2008 An Engineered Protein Tag for Multiprotein Labeling in Living Cells *Chemistry & Biology* **15** 128–36

[163] Los G V, Encell L P, McDougall M G, Hartzell D D, Karassina N, Zimprich C, Wood M G, Learish R, Ohana R F, Urh M, Simpson D, Mendez J, Zimmerman K, Otto P, Vidugiris G, Zhu J, Darzins A, Klaubert D H, Bulleit R F and Wood K V 2008 HaloTag: A Novel Protein Labeling Technology for Cell Imaging and Protein Analysis *ACS Chem. Biol.* **3** 373–82

[164] Klein T, Löschberger A, Proppert S, Wolter S, van de Linde S and Sauer M 2011 Live-cell dSTORM with SNAP-tag fusion proteins *Nature Methods* **8** 7–9